\documentclass{JHEP3}
\usepackage{epsfig}
\usepackage{amsmath}
\usepackage{mathrsfs}

\newcommand{\be}{\begin{equation}}
\newcommand{\ee}{\end{equation}}
\newcommand{\ba}{\begin{eqnarray}}
\newcommand{\ea}{\end{eqnarray}}

\title{Group theoretical approach to \\ quantum fields in de Sitter space \\ I. The principal series}

\author{E. Joung$^{ab}$, J. Mourad$^{ab}$ and R. Parentani$^b$\\
	$^a$Astro Particules et Cosmologie,\footnote{Unit\'e Mixte de Recherche du CNRS (UMR 7164).}
	Universit\'e  Paris VII,\\
	2 place Jussieu - 75251 Paris Cedex 05, France\\
	$^b$Laboratoire de Physique Th\'eorique,\footnote{Unit\'e Mixte de Recherche du CNRS (UMR 8627).}
	B\^at. 210, Universit\'e Paris XI,\\ 
	91405 Orsay Cedex, France\\
	E-mail: \email{joung@th.u-psud.fr}, \email{mourad@th.u-psud.fr}, \email{parenta@th.u-psud.fr}}

\abstract{Using unitary irreducible representations of the de Sitter group,
 we construct the Fock space of a massive free scalar field.
 In this approach, the vacuum is the unique dS invariant state. The quantum field is
\emph{a posteriori} defined by an operator subject to covariant
transformations under the dS isometry group. This
insures that it obeys canonical commutation relations, up to an overall factor
which should not vanish as it fixes the value of $\hbar$. However, contrary to what is
obtained for the Poincar\'e group,
the covariance condition leaves an arbitrariness in the definition of the field. This
arbitrariness  allows to recover the amplitudes
governing spontaneous pair creation processes, as well as the
class of alpha vacua obtained in the usual field theoretical
approach. The two approaches can be  formally related by introducing a
squeezing operator which acts on the state in the field
theoretical  description and on the operator
in the present treatment. The choice of the different dS invariant schemes
(different alpha vacua) is here posed in very simple terms:
it is related to a first order differential equation which is
singular on the horizon and whose general solution is therefore
characterized by the amplitude on either side of the horizon.  Our algebraic approach
offers a new method to define quantum field theory
on  some  deformations of dS space.}


\begin{document}

\section{Introduction}

In flat spacetime, there are several equivalent ways to quantize a
massive free scalar field.
One can start with the invariant action, impose the canonical commutation relations, and
derive the generators of the Poincar\'e group as Noether charges \cite{itz}.
From a Hamiltonian point of view, one has a collection of harmonic
oscillators labelled by the spatial momentum.
 One then verifies that the one-particle
sector of the Fock space of these oscillators is a unitary 
irreducible representation (UIR)
of the Poincar\'e group \cite{wig}.

Alternatively, one can start from the UIR
of the Poincar\'e group and construct  a Fock space as the direct
sum of the symmetrized tensor product of the UIR Hilbert space 
\cite{Weinberg:1995mt, Weinberg:1996kw}.
One can then introduce creation and annihilation operators
which are associated to a basis in the UIR Hilbert space, and
which are linearly combined to form a local field. The latter is
uniquely defined by the requirement that it should covariantly
transform:
\be
   \Phi(\Lambda x)=U(\Lambda)\, \Phi(x)\, U^\dagger(\Lambda)\,,
   \label{e1}
\ee
where $U(\Lambda)$ is the unitary operator representing the Poincar\'e
transformation $\Lambda$ in the  Fock space. This procedure is explained in
more detail in Section 2.

Thirdly, one can quantize a massive scalar field
 from an analysis of Green functions $G(x;x')$, invariant solutions of
the Klein-Gordon equation with well defined boundary conditions.
Explicitely, for the positive frequency Wightman function, one writes
\be
   G_+(x;x')=\sum_{n}\phi_n(x)\, \phi_n^*(x')\,,
   \label{posw}
\ee
where $\{\phi_n,\phi_n^*\}$ form a complete  set of orthogonal solutions
(of norm $+1, -1$)  with respect to the Klein-Gordon product.
One can then define the field operator as
\be
   \Phi(x)=\sum_n\phi_n(x)\, a_n+\phi_n^*(x)\,a^\dagger_n\,.
\ee
The equal time canonical commutation relations
 are then equivalent to
\be
   [\,a_n,\,a^\dagger_{n'}\,]=\delta_{n,n'}\,, \qquad [\,a_n,\, a_{n'}\,] = 0\,.
\ee
In this approach the modes $\phi_n$ are univocally defined by
the positive frequency condition implied by eq.(\ref{posw}).

When quantizing fields in dS space, one encounters
novel features which engender difficulties
\cite{Chernikov:1968zm,ge,Tagirov:1972vv,Bunch:1978yq,
Gibbons:1977mu,Mottola:1984ar,Allen:1985ux,Allen:1987tz}
and which are deeply rooted in the absence of a time-like Killing vector field
which is globally defined, see Figure \ref{fig1}.
 \EPSFIGURE{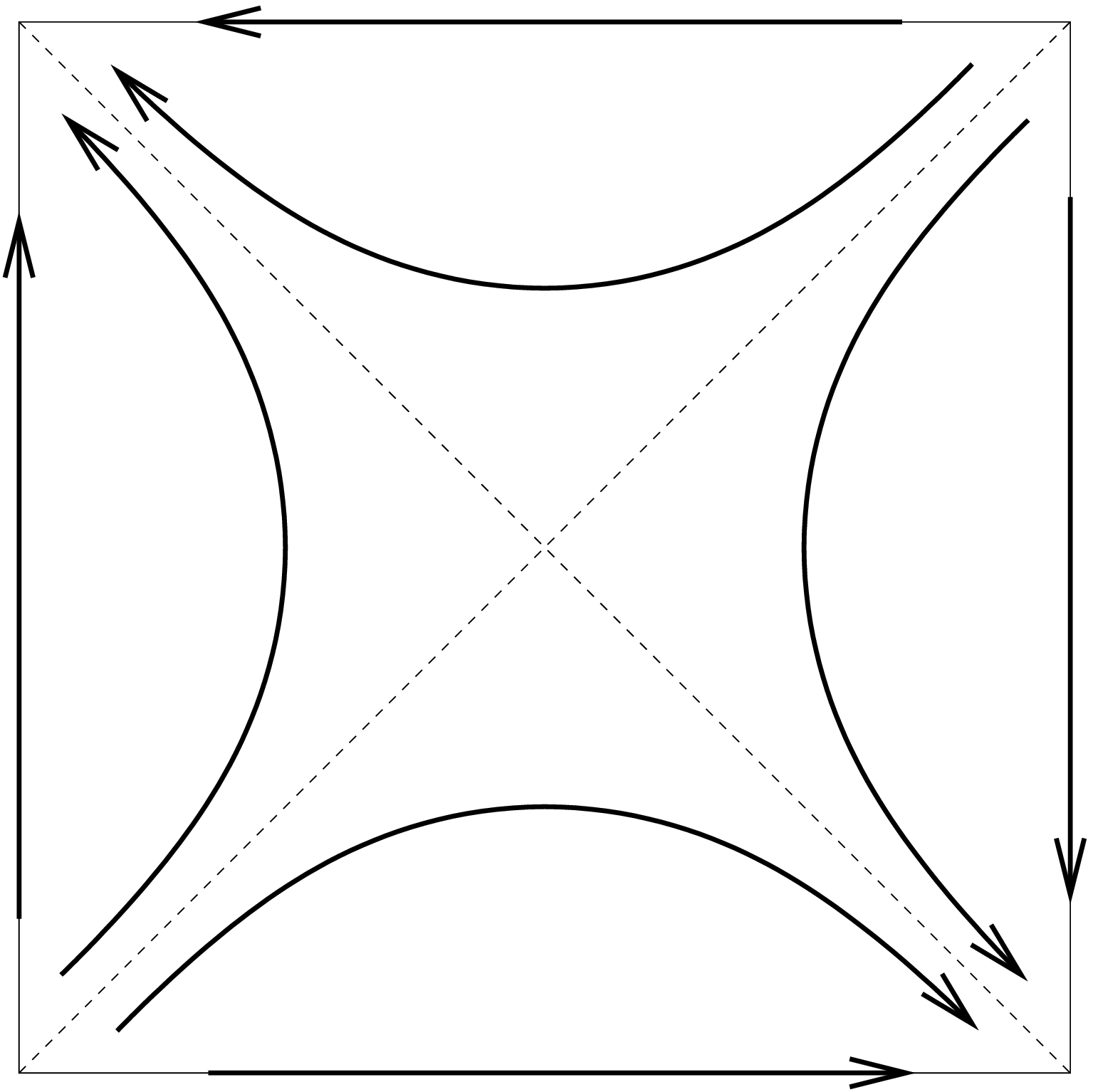,width=5cm}{\label{fig1}
 The flow of a boost Killing vector field is represented in 
a Carter-Penrose diagram of de Sitter space. The bifurcating horizon is the locus
where the null horizons meet. The Killing vector field is time-like and future directed 
only in the left quadrant, i.e. for the points which are space-likely separated 
from the bifurcating horizon, and on its left. The vector fields of the
other isometries of dS space are all space-like.}
The approach which has been mostly followed is  the third  one we just
mentioned. It has been realized that there exists a
two-parameter family of dS invariant Green functions (i.e.
functions of a quantity which coincides with the dS invariant
geodesic distance whenever the latter exists between the two
points under consideration).

The physical interpretation of the degeneracy is
provided in terms of different vacuum states related to each other
by Bogoliubov transformations \cite{Allen:1985ux}. These 
states are generally referred as \emph{alpha vacua}.

An additional important notion is provided by the behavior of the Green
function in the coincidence point limit.  When requiring that this
behavior be that of Hadamard \cite{birelldavies}, which is a covariant way to impose vacuum
condition in the flat space-time limit, only one alpha vacuum is picked out.  It
coincides with the so-called Bunch-Davies (BD) vacuum defined by a
positive conformal frequency condition in the high momentum limit \cite{Bunch:1978yq}.
There are two interesting aspects related to this state which are
worth mentioning. First, the BD vacuum is perceived by an inertial
particle detector as a thermal bath at a temperature given by
$H/2\pi$ where $H$ is the Hubble parameter. This is in agreement
with the dS horizon temperature derived from the first law of
thermodynamics and the Bekenstein-Hawking entropy \cite{Gibbons:1977mu} associated to the area
of the cosmological horizon.
The second aspect arises from the fact the initial vacuum, conventionely
defined by the absence of massive quanta as seen by an inertial
particle detector at asymptotic early times, neither coincides with the BD
vacuum nor does it with the final vacuum \cite{Mottola:1984ar}. This vacuum instability
results from the spontaneous creation of pairs of massive
particles in dS space.  One thus faces a dilemma: either one
insists on the Hadamard character and one works with the BD vacuum
which contains coherent superpositions of pairs of particles at
early times, i.e. a state which cannot be prepared at early times,
 or one gives up the Hadamard condition and one deals
with a state whose energy density is infinite using the standard
renormalization procedure. The only way to get states which are both
Hadamard and contain no particle at early times is to break the dS invariance
 of the state.

In this paper, we reconsider the quantization of a massive scalar
field from a group theory point of view. Our approach is close to the
second method in the above list. Namely, we start with the UIR of
dS group \cite{rep1}-\cite{vil}, that is, SO$(1,n)$ where $n$ is the space-time dimension. The
representations  fall into three different classes called the
principal, the complementary, and the discrete series \cite{vil}. Each series
requires a separate analysis. In this paper we study the principal series, the other two series will be presented in
a separate paper.

The representations belonging to the  principal series are univocally
given in terms of the UIR of the \emph{compact} group SO$(n-1)$ and a real parameter
which fixes the quadratic Casimir together with
the mass of the field. From the Hilbert space of this UIR and the trivial
representation, we  build a Fock space wherein there is
a unique dS invariant vacuum state.
In this construction therefore,  we  do not recover the
above mentioned two-parameter ambiguity of the dS invariant vacua.
In fact, it is our aim to discover how will this
ambiguity reappear when constructing the local field from the UIR.
It is also our endeavor to understand how the notion of vacuum instability
translates in terms of group representations.
Both questions are explicitly answered by the construction we present.

To construct the Fock space we start with the trivial
representation and the Hilbert space carrying the UIR. Explicitly,
the vacuum corresponds to the trivial representation, the one
particle sector to the UIR, and the $n$ particles sectors are
given by  symmetrized tensor products of the latter Hilbert space.
We can then define creation and annihilation operators, and
realize the generators of the dS group on this Fock space as
bilinear operators in $a$ and $a^\dagger$. This structure
guarantees that each sector with a given particle number remains
invariant under dS transformations. Moreover if interactions
among sectors are expressed in terms of these operators there is
still no ambiguity in the quantum theory.
In fact it is only through the requirement that the interactions be local in space-time
that the ambiguity reappears. Indeed, when constructing 
local field operators from the UIR,
we get a family of covariant and canonical 
fields parameterized by elements of SU$(1,1)/$U$(1)$. Two members of
the family are formally related by an infinite product of  (two-mode) squeezing operators which
commutes with all generators of dS group.
Contact with the usual treatment is made by computing the Green functions
associated with the various field operators. In a nutshell one has
the following equalities:
\ba
    G_{\alpha,\beta}(x,x')&=&\langle
   \Omega|\,\Phi_{\alpha,\beta}(x)\,\Phi_{\alpha,\beta}(x')\,|\Omega\rangle\nonumber\\
   &=&\langle
   \Omega|\,{\mathcal{S}}_{\alpha,\beta}^\dagger\,\Phi(x)\,\Phi(x')\,{\mathcal{S}}_{\alpha,\beta}
   |\Omega\rangle\nonumber\\
   &=&\langle \Omega_{\alpha,\beta}|\,\Phi(x)\,\Phi(x')\,
   |\Omega_{\alpha,\beta}\rangle\,,
\ea
where $|\Omega\rangle$ is the unique vacuum
state, $\Phi_{\alpha,\beta}$ is the local field labelled by $(\alpha,\beta)$,
which are coordinates of SU$(1,1)/$U$(1)$, and ${\mathcal{S}}_{\alpha,\beta}$
is the squeezing operator relating this field
to a reference one, e.g. characterized by $(0,0)$.
In the third line $|\Omega_{\alpha,\beta}\rangle$ is the
corresponding alpha vacuum as usually defined. The correspondence between the
two treatments is made unambiguous by the
identifying the Green function associated with the BD vacuum in both of them.

Besides offering an interesting alternative way to recover known results,
the present approach possesses several advantages with respect to the
usual approach.
For instance, the various solutions for the field operator
arise from a first order differential equation
which is singular on the cosmological horizon.
This offers a simple and geometrical interpretation of
the solutions in terms of the field amplitude on either side of the
horizon. In particular it offers a straightforward
identification of the three most relevant cases,
the BD, the initial and final vacua. In this respect,
we point out that the quantization
of a scalar field in a constant electric field is also efficiently obtained
by a similar treatment based on a first order differential equation
which is singular on the (acceleration) horizon \cite{ParBroutNPB,PhysRep95}. 
In that case as well,
the amplitudes of the solutions on either side directly govern the
pair creation probability amplitudes.\footnote{
The possibility of re-expressing, in the presence of Killing
horizons, second order differential equations
by singular first order equations which  encode the holomorphic
character of the positive frequency modes
in an interesting feature which deserves further study as it
leads to horizons thermodynamics \cite{JacP} and encompasses
in a unified treatment, electro-production, Hawking effect,
Unruh effect, and pair creation in de Sitter space.}

An additional advantage of the group theory approach is that
 explicit solutions of the field operator are obtained
 from the first order equation
 without having to solve the Klein-Gordon equation.
In a similar vein, Green functions are given by well defined integrals.
In addition,  these  include the $i\epsilon$ prescription encoding
the holomorphic properties of the modes on the horizon.

Moreover, given that the isometry groups and therefore the representations
of dS and AdS in (1+1)-dimensions
are identical (modulo the identification of the mass square),
the local field we constructed can be viewed as living on AdS
upon exchanging  temporal and  spatial coordinates and momenta.
The principal series here considered describes tachyonic scalar fields
in AdS.

Finally, we hope that the  present algebraic approach to quantum field
theory on dS space will offer interesting applications when
considering deformations of the dS group, which could result from
some noncommutative description of spacetime \cite{MJ} or perhaps also from
self gravitational effects in (2+1)-dimensions \cite{Buf}.  In particular, the
calculation of the modified pair creation amplitudes could proceed
along the same lines as the ones we adopted.

The paper is organized as follows. In Section 2, we present
our method by considering the quantization of a massive scalar
field in a flat two-dimensional spacetime. We review in Section 3 the
basic properties of the representations belonging to the principal
series of the dS group in two dimensions. In Section 4, we construct the
local field, verify that it automatically obeys canonical
commutation relations and study the effect of time reversal. In
Section 5 we compute the Wightman function. This allows to
identify the particular field giving rise to the BD vacuum. All the other
solutions can then be expressed in terms of elements of
SU$(1,1)/$U$(1)$. This offers a direct comparison with the
parameterization of the Green functions given by Allen \cite{Allen:1985ux} and
therefore to make contact with the class of alpha vacua. To complete the
quantization we study, in Section 6,
 the time evolution of the field operator, we identify the initial and final
 positive frequency modes, and compute the pair creation
 probability amplitudes.  In Section 7, we construct the squeezing
 operator relating couples of alpha vacua in the usual approach, and  pairs of canonical  fields
 in our treatment. Finally, in Section 8 we generalize this algebraic approach to
 the $n$-dimensional case and show that the moduli space of canonical fields remains SU$(1,1)/$U$(1)$.
 In Appendix A, we compute the Wightman functions in terms of the holomorphic and anti-holomorphic
 solutions of the above mentioned first order differential equation.
In Appendix B,  we use our algebraic formulation  to describe the quantization
 of the scalar field defined on the flat sections of dS space.

\section{Massive scalar field in Minkowski spacetime}

In order to illustrate the method we shall use, let us first consider
a massive scalar field in the two-dimensional flat space-time. We shall show how
starting from a Unitary Irreducible Representation (UIR) of the
Poincar\'e group, we can construct the corresponding
canonical scalar field operator.

The Poincar\'e group is generated by a space translation operator,
 a time translation and a boost. When these operators act on the UIR,
they are respectively noted by $\cal P$, $\cal H$ and $\cal K$.
The algebra is
\be
    [\,{\cal P},\, {\cal H}\,]= 0\,,\qquad[\,{\cal K},\, {\cal P}\,] 
    = i {\cal H}\,,\qquad [\,{\cal K},\, {\cal H}\,] = -i {\cal P}\,.
\ee
The irreducible scalar
representations of the Poincar\'e group are characterized
by the value of the quadratic Casimir ${\cal C} = -{\cal H}^2 + {\cal P}^2 = M^2$
and the sign of the energy. At
fixed $M$, the representation is realized on the Hilbert space
${\mathscr H}$ with basis $|{\bf{p}}\rangle$, and scalar product
$\langle {\bf{p}}'|{\bf{p}}\rangle=\delta({\bf{p}}-{\bf{p}}')$ as
\ba
   {\mathcal P}\,|{\bf{p}}\rangle &=&{\bf{p}}\,|{\bf{p}}\rangle\,,
   \qquad {\cal H}\,|{\bf{p}}\rangle =\omega({\bf{p}})\,|{\bf{p}}\rangle\,, \nonumber\\
   {\cal K}\,|{\bf{p}}\rangle &=&{i\over 2}\left\{\frac{\partial}{\partial{\bf{p}}},
   \,\omega({\bf{p}})\right\}\,|{\bf{p}}\rangle\,,
\ea
where $\omega({\bf{p}})=\sqrt{{\bf{p}}^2+M^2}$
is the positive root of ${\cal C}= M^2$.

From the Hilbert space ${\mathscr H}$ and the trivial
representation of the Poincar\'e group
of Hilbert space ${\mathscr H}_0$ (of dimension one and spanned by the vector $\vert\Omega\rangle$),
we construct the Fock space  as
\be
     {\mathscr F}= {\mathscr H}_0 \oplus
    \bigoplus_{n=1}^{\infty}{\bigotimes_s^n}{\mathscr H}\,,
\ee
where the $n^{\rm th}$
term is the representation obtained by taking the symmetrical
tensor product of $n$ copies of the UIR. The Fock space carries a
reducible unitary representation of the Poincar\'e group. The generators will be noted
$P$, $H$ and $K$, and the group elements $U(\Lambda)$. We also introduce creation and
annihilation operators and denote them $a^\dagger({\bf{p}})$ and $a({\bf{p}})$.
They obey
\ba
   a^\dagger({\bf{p}}) \vert \Omega \rangle &=& \vert {\bf{p}} \rangle\,,
   \nonumber\\
   \left[\,a({\bf{p}}),\, a^\dagger({\bf{p}}') \,\right] &=&\delta({\bf{p}}-{\bf{p}}')\,,
   \label{aop}
\ea
where $\vert\Omega\rangle$  is annihilated by all the
annihilation operators $a({\bf{p}})$ and is thus the vacuum of the Fock space.
More explicitely the operators acting on Fock space are
\be
    U(\Lambda)=\int d{\bf p} d{\bf p'}\,\langle{\bf p}|\,{\cal U}(\Lambda)\,
    |{\bf p'}\rangle\, a^\dagger({\bf p})\,a({\bf p}')\,,
\ee
where ${\cal U}(\Lambda)$ acts on the UIR.

Our goal now is to construct a \emph{local} field operator
 on ${\mathscr F}$, $\Phi(x)$, which is a linear superposition
of creation and annihilation operators.
The basic requirement on this local field is the covariance
transformation property eq.(\ref{e1}).
This implies that the field is determined by
its value at the origin:
\be
   \label{e1p}
   \Phi(x)=e^{-ix\cdot P}\, \Phi(0)\, e^{ix\cdot P}\,,
\ee
where $x\cdot P=-tH+{\bf x}P$. It also implies that $\Phi(0)$ is invariant under the boost transformation 
which leaves the origin fixed:
\be
   [\,K,\,\Phi(0)\,]=0\,.
   \label{b}
\ee

Writing $\Phi(0)$ as
\be
   \Phi(0)=\int d{\bf{p}}
   \,\Big(\, c({\bf{p}})a^\dagger({\bf{p}})+c^*({\bf{p}})a({\bf{p}})\,\Big)\,,
\ee 
eq.(\ref{b}) implies that 
\be
   \omega({\bf{p}})c'({\bf{p}})+\frac12c({\bf{p}})\omega'({\bf{p}})=0\,,
\ee
which gives
\be
   c({\bf{p}})={A \over \sqrt{\omega({\bf{p}})}}\,,
\ee
where $A$ is a constant complex amplitude.
It should be noticed that $c({\bf p})$ is even in $\bf p$.
This implies that $\Phi(0)$ is left invariant under  parity.
It is also interesting to study the behavior of $\Phi(0)$ under
time reversal $t \to-t$ which is the other global symmetry
which leaves $(0,0)$ invariant.
It is implemented by the (anti-unitary) operator satisfying
\be
    {\mathsf T}\,H\,{\mathsf T}^{-1}=H\,,\qquad
    {\mathsf T}\,P\,{\mathsf T}^{-1}=-P\,,
    \qquad {\mathsf T}\,K\,{\mathsf T}^{-1}=K\,.
\ee
When acting on the field operator one has
\be
    {\mathsf T}\,\Phi_{\{A\}}(t,{\bf x})\,{\mathsf T}^{-1}=\Phi_{\{A^*\}}(-t,{\bf x})\,,
\ee
where we have  written $A$ as a subscript to
make explicit that the field depends on it.

It should be emphasized that the linear scalar field $\Phi$, the solution of eq.(\ref{e1}) 
automatically
satisfies the equal time commutators. Indeed, one has 
\be
    [\,\Phi(0,0),\,\Phi(0,{\bf x})\,] = 0\,,
\ee
and
\be
   [\,\partial_t\Phi(0,0),\,\Phi(0,{\bf x})\,]
   =-4\pi i{|A|^2}\delta({\bf x})\,.
\ee
The latter has the required $\delta$ dependence, and
can thus be used to fix the norm of $A$ in terms of $\hbar$, namely $4\pi |A|^2= \hbar$.
(The phase of $A$ can be
absorbed in the definition of the operators $a({\bf{p}})$).
Therefore the operator
\be
   \Phi(x)=\sqrt{\hbar\over 4\pi}\int {d{\bf{p}}
   \over{\sqrt{\omega({\bf{p}})}}} \Big(\,a({\bf{p}})\,
   e^{ix\cdot p}\,+\,a^\dagger\, ({\bf{p}})\,e^{-ix\cdot p}\,\Big)\,.
\ee
is a canonical scalar  field. It should be also noticed that
\be
    \Delta(x)=[\,\Phi(x),\,\Phi(0)\,]=2i\int d{\bf{p}}
    \,|c({\bf{p}})|^2\,\sin(x\cdot p)\,.
\ee 
vanishes for causally disconnected points. The above field is therefore also causal, 
by the construction we adopt. 

In conclusion, we found that, up to an irrelevant phase (that of $A$), 
$\Phi$ is univocally defined by eq.(\ref{e1}).
This will not be true for the de Sitter group, as we now show.

\section{The SO$(1,2)$ group}

The starting point of our approach is the UIR of the SO$_0(1,n)$
group, the group of linear transformations with determinant $1$
which leaves $-(X^0)^2+(X^1)^2+\dots +(X^n)^2$ invariant
and which are connected to the identity. This is the isometry group
of $dS_n$, the de Sitter space with $n$ space-time
dimensions. The UIR were first analyzed  by Bargmann \cite{rep1}
for $n=2$, Gelfand and Naimark for $n=3$ \cite{gel}, Thomas, Newton and
Dixmier for $n=4$ \cite{th,rep2}. General aspects for all $n$ were studied in 
\cite{rep3,rep4,rep5,rep6}.
In this paper we shall be concerned with the principal series
representations.\footnote{In a contraction  limit \cite{cont},
one can relate
 these representations of the dS group to those of
 the Poincar\'e group. All
massive representations of the Poincar\'e group can be obtained
this way \cite{rep5}. However, the resulting representations are not
irreducible, since one obtains a product of
a positive and a negative energy UIR.
This doubling is directly related to the
afore mentioned two-parameter ambiguity in constructing local
fields from UIR of the dS group.}

In the following we shall concentrate on the two-dimensional de
Sitter space with group  SO$_0(1,2)$. Let $\cal J$ be the generator of
the rotation subgroup and ${\cal K}_1$ and ${\cal K}_2$ the two boosts. They
verify the commutation relations:
\be
   [\,{\cal J,\, K}_1\,]=i{\cal K}_2\,,\qquad [\,{\cal J,\,K}_2\,]=
   -i{\cal K}_1\,,\qquad [\,{\cal K}_1,\,{\cal K}_2\,]=-i{\cal J}\,.
\ee
The quadratic Casimir
operator
\be
   {\cal C=J}^2-{\cal K}_1^2-{\cal K}_2^2\,,
\ee
commutes with all the
generators and is constant on an irreducible representation.
Bargmann classified the UIR according to the value of $\cal C$ and the
eigenvalues $m$ of $\cal J$:
\begin{itemize}
   \item
       (i) the principal series with ${\cal C}\le-{1\over4}$,\quad
       $m=0,\pm 1,\dots$ or $m=\pm{1\over 2},\pm{3\over 2},\dots$;
   \item
       (ii) the complementary series with $-{1\over 4}<{\cal C}<0$ and $m=0,\pm1,\dots$;
   \item
       (iii) the discrete series $D^+_k$ with $k$
       non-negative integer or half integer,
       ${\cal C}=k(k+1)$ and $m=k+1,k+2,\dots$ and finally
   \item
       (iv) the discrete series
       $D^-_k$,  ${\cal C}=k(k+1)$ and $m=-(k+1),-(k+2),\dots$
\end{itemize}

For the representations of the principal series, the value of the Casimir operator
is given by $-\left(\mu^2+{1/4}\right)$ with real $\mu$. 
At fixed $\mu$ the generators of SO$_0(1,2)$ act on
the basis of the normalized eigenstate of ${\cal J}$ as
\be
   {\cal J}|m\rangle=m\,|m\rangle\,,\qquad
   {\cal K}_{\pm}|m\rangle=\left[\pm i\,m
   -i\left(i\mu-{1\over 2}\right)\right]|m \pm 1\rangle\,.
\ee
where the raising and lowering operators are defined as ${\cal K}_{\pm}={\cal K}_1\pm i{\cal K}_2$.

The  representations with integer $m$ can be also
realized on the Hilbert space of square integrable complex (univalued) functions on
the circle, $\Psi(\phi)$, equipped with the standard ${\cal L}^2$ scalar product.
We shall use the Dirac notation where $\Psi(\phi)=\langle\phi\vert\Psi\rangle$.

The action of the group generators has the following expressions:
\ba
   \langle\phi\vert {\cal J}\vert\Psi\rangle &=&-i{d \over
   d\phi}\langle\phi\vert\Psi\rangle\,
   ,\nonumber \\
   \langle\phi\vert {\cal K}_1\vert\Psi\rangle
   &=&\left[ i\sin\phi{d \over d\phi}-i\left(i\mu-{1\over 2}\right)\cos\phi\right]\langle\phi\vert\Psi\rangle
   =\left[{i\over 2}\left\{\sin\phi,{d \over d\phi}\right\}+\mu\cos\phi\right]\langle\phi\vert\Psi\rangle\,,
   \nonumber\\
   \langle\phi\vert {\cal K}_2\vert\Psi\rangle
   &=&\left[-i\cos\phi{d \over d\phi}-i\left(i\mu-{1\over 2}\right)\sin\phi\right]\langle\phi\vert\Psi\rangle
   =\left[-{i\over 2}\left\{\cos\phi,{d \over d\phi}\right\}+\mu\sin\phi\right]\langle\phi\vert\Psi\rangle\,.
   \nonumber\\
\ea
They are hermitian if $\mu$ is real.

It will be usefull to also  have the action of finite
transformations. They are given by
\ba
   \langle\phi\vert \,e^{i\theta {\cal J}}\vert \Psi\rangle
   &=& \langle\phi+\theta\vert\Psi\rangle\,,
   \nonumber\\
   \langle\phi\vert \,e^{i\rho {\cal K}_1}\vert\Psi\rangle
   &=&(\cosh\rho +\sinh\rho\cos\phi)^{i\mu-1/2} \,\langle\phi_1\vert\Psi\rangle\,,
   \nonumber \label{te}\\
   \langle\phi\vert \,e^{i\lambda {\cal K}_2}\vert\Psi\rangle
   &=& (\cosh\lambda +\sinh\lambda\sin\phi)^{i\mu-1/2}\, \langle\phi_2\vert\Psi\rangle\,.
\ea
where
\ba
   \cos\phi_1 &=& {\cos\phi  \cosh\rho+\sinh\rho\over \cosh\rho+\sinh\rho \cos\phi}\,,\qquad
   \sin\phi_1={\sin\phi \over \cosh\rho +\sinh\rho\ \cos\phi}\,,
   \nonumber\\
   \cos\phi_2&=&{\cos\phi
   \over \cosh\lambda +\sinh\lambda\ \sin\phi}\,,\qquad \sin\phi_2={\sin\phi \cosh\lambda+\sinh\lambda
   \over \cosh\lambda +\sinh\lambda\ \sin\phi}\,.
\ea

\section{Massive scalar field in de Sitter space}

Let $|\Omega\rangle$ be a state carrying the trivial representation and $\mathscr H$ the Hilbert space carrying
the UIR of the preceeding section. On the Fock space constructed from these two subspaces we define the creation
and annihilation operators $a_m$ and $a^\dagger_m$ as in eq.(\ref{aop}).

The (reducible) representation of the dS group on the Fock space is deduced
from the irreducible representation by
\be
  U=\sum_{m, m'} {\cal U}_{mm'} \, a^\dagger_m\,a_{m'}\,,
\ee
where ${\cal U}_{mm'}=\langle m|\,{\cal U}\,|m'\rangle$. Each
$n$-particle sector of the Fock space is thus kept invariant under the
action of the group transformations.

Using these operators, we shall now construct a local field on
dS space: $\Phi(x)$. We shall use  the \emph{global} coordinate system $(t,\theta)$, where
$t$ is the time coordinate and varies from $-\infty$ to $+\infty$
and $\theta$ is an angle coordinate. In this coordinate system, the metric is
\be
   ds^2=-dt^2+\cosh ^2t\,d\theta^2\,.
\ee
The de Sitter space can be described by its
embedding in a flat three dimensional space $-(X^0)^2+(X^1)^2+(X^2)^2=1$ with
\ba
   X^0&=&\sinh t\,,\nonumber\\
   X^1&=&\cosh t\cos\theta\,,\nonumber\\
   X^2&=&\cosh t\sin\theta\,.
\ea
The point with global coordinates $(0,0)$ is transported to
$(t,\theta)$  by the following \emph{ordered} sequence: first one acts with
a boost generated by $K_1$ with parameter $t$ to reach $(t,0)$, and
then one reaches $(t,\theta)$ by a rotation with angle $\theta$. Notice also
that the point $(0,0)$ is left invariant by the boost generated by $K_2$.

As in eq.(\ref{e1p}), the covariance condition (\ref{e1}) implies that $\Phi(t,\theta)$ can be deduced from
$\Phi(0,0)$.
 Taking into account the non-commuting character of $J$ and $K_1$ we have
 the central equation:
\be
   \Phi(t,\theta)=e^{-iJ\theta}e^{itK_1} \, \Phi(0,0) \, e^{-itK_1}e^{iJ\theta}\,.
   \label{e1dS}
\ee
We have adopted a sign convention such that the expanding
phase of dS space corresponds to positive $t$. As a direct
consequence we have
\ba
   -i\partial_t \Phi(t,\theta)&=&[\,K_1\cos\theta+K_2\sin\theta,\,\Phi(t,\theta)\,]\,,\label{ev1}\\
   i\partial_\theta\Phi(t,\theta)&=&[\,J,\,\Phi(t,\theta)\,]\,.\label{ev2}
\ea

As in Minkowski space, the field operator
$\Phi(0,0)$ must be invariant under transformations that leave the point $(0,0)$
invariant. From the dS group only $K_2$ leaves $(0,0)$ invariant. Hence we have
\be
   \label{e1dS2}
   [\,K_2,\,\Phi(0,0)\,]=0\,.
\ee 
Using eq.(\ref{e1dS}) this condition leads to 
\be
   [\,\cosh t\,(\cos\theta K_2-\sin\theta K_1)+\sinh t J,\,\Phi(t,\theta)\,]=0\,.
   \label{ev3}
\ee 
Using eq.(\ref{ev1}-\ref{ev3}), we get 
\ba
   \left[\,K_1,\,\Phi(t,\theta)\,\right]&=&i\,(\tanh t\sin\theta \partial_\theta-\cos\theta\partial_t)
   \Phi(t,\theta)\,,
   \nonumber\\
   \left[\,K_2,\,\Phi(t,\theta)\,\right]&=&-i\,
   (\tanh t\cos\theta \partial_\theta+\sin\theta\partial_t)
   \Phi(t,\theta)\,.
   \label{is}
\ea
The right hand sides of the above equations coincide as they should with the
Killing vector fields on dS space acting on the scalar field.
The Casimir relation for the group generators gives rise to
\be
   \Big[J,\left[J,\Phi\right]\Big]-\Big[K_1,\left[K_1,\Phi\right]\Big]-
   \Big[K_2,\left[K_2,\Phi\right]\Big]=
   -\left(\mu^2+{1\over 4}\right)\Phi\,.
\ee 
Upon using eq.(\ref{is}) one verifies that $\Phi(t,\theta)$ obeys the Klein-Gordon equation:
\be
   \left(\partial_t^2+\tanh t\,\partial_t-\frac1{\cosh ^2t}\,\partial_\theta^2\right)
   \Phi(t,\theta)=-\left(\mu^2+{1\over 4}\right)\Phi(t,\theta)\,.
\ee
This allows the identification
of the mass squared in terms of $\mu$:
$M^2+\xi R=H^2(\mu^2+{1\over 4})$,
where $\xi$ is, in the action formalism, the coefficient of the term
in $\frac12R \Phi^2$.

We now search for solutions of eq.(\ref{e1dS2}) which are linear in the creation and annihilation operators.
This is our second assumption. We start in the Fourier basis and expand
 $\Phi(0,0)$ as
\be
   \Phi(0,0)=\sum_{m=-\infty}^{\infty}\ c_m
   a^\dagger_m + c^*_m a_m\,.
\ee 
The field operator is thus fully determined by the c-number constants $c_m$. The covariance condition
implies eq.(\ref{e1dS2}), which in turn
 determines the constants
$c_m$ as we now show. From
\be
   [\,K_2,\,a_m^\dagger\,]={1\over 2}\left[\left( m -\Big(i\mu-{1\over
   2}\Big)\right)a^\dagger_{m+1} +\left( m +\Big(i\mu-{1\over
   2}\Big)\right)a^\dagger_{m-1}\right]\,,
\ee
we get
\be
   c_{m-1}\left( m-{1\over
   2} -i\mu\right)= -c_{m+1}\left( m+{1\over 2}
   +i\mu\right)\,.
   \label{inva}
\ee
This equation determines the constants
$c_m$ in terms of $c_0$  and $c_1$ for $m$ even and odd respectively.
More explicitly the above
relation gives
\be
   {c_{m+2}\over{c_m}}={\gamma_{m+2}\over\gamma_m}\,,
\ee
with
\be
   \gamma_m=e^{im\pi\over2} {\Gamma\left({m\over 2}+{1\over 4}-{i\mu
   \over 2}\right) \over \Gamma\left({m\over 2}+{3\over 4}+{i\mu
   \over 2}\right)}\,,
   \label{gam}
\ee
which implies
\be
   c_{2m}={c_0\over
   \gamma_0}\gamma_{2m},\quad c_{2m+1}={c_1\over
   \gamma_1}\gamma_{2m+1}\,.
\ee

Notice that the coefficients $c_m$ are even:  $c_{m}=c_{-m}$, as for the Poincar\'e group.
This guarantees the invariance under the parity transformation $\mathsf{P}$.
Parity acts on the UIR as ${\mathsf{P}}|m\rangle=  \vert -m\rangle$
and on the field as ${\mathsf{P}}\,\Phi(t,\theta)\,{\mathsf{P}}^{-1}=\Phi(t,-\theta)$.

For de Sitter group it is also  instructive to work out the arbitrariness in
the expansion of $\Phi(0,0)$ in the position basis $|\phi\rangle$.
 We find convenient to present the analysis in terms of states in the UIR
 rather than field operators.
 We shall therefore study the states $\vert \Psi_0\rangle=
\Phi(0,0)|\Omega\rangle$
which are invariant under $K_2$, namely they must obey 
$K_2 \vert \Psi_0 \rangle =0$.
Conversely an invariant state defines
coefficients $c_m$ obeying the above equations.
In the $\phi$ representation the equation $\langle \phi \vert K_2 \vert \Psi_0 \rangle =0$
gives rise to a singular first order equation:
\be
   \cos\phi{d\Psi_0(\phi) \over d\phi}+\left(i\mu-{1\over
   2}\right)\sin\phi\,\Psi_0(\phi)=0\,.
   \label{k2}
\ee
Its solution is given by
\be
\Psi_0(\phi)=\left\{
   \begin{array}{ll}
       A \, (\cos\phi)^{i\mu-{1\over 2}}, & {\rm for} \ -{\pi\over 2}<\phi<{\pi\over 2}\,,\\
       &\\
       B \, (-\cos\phi)^{i\mu-{1\over 2}}, & {\rm for}\quad {\pi\over 2}<\phi<{3\pi\over 2}\,.
   \end{array}
   \right.
\ee
It is important to notice that although the equation is first order,
being singular at $\phi=\pm {\pi\over 2}$, its solution
depends on two complex numbers.  The region
$-{\pi\over 2}<\phi<{\pi\over 2}$ is the spatial region
in the causal past of 
$(0,0)$: if a signal is emitted from this region at a sufficiently early time
it can reach the space-time point (0,0) as can be seen in Figure \ref{fig2}.
\EPSFIGURE{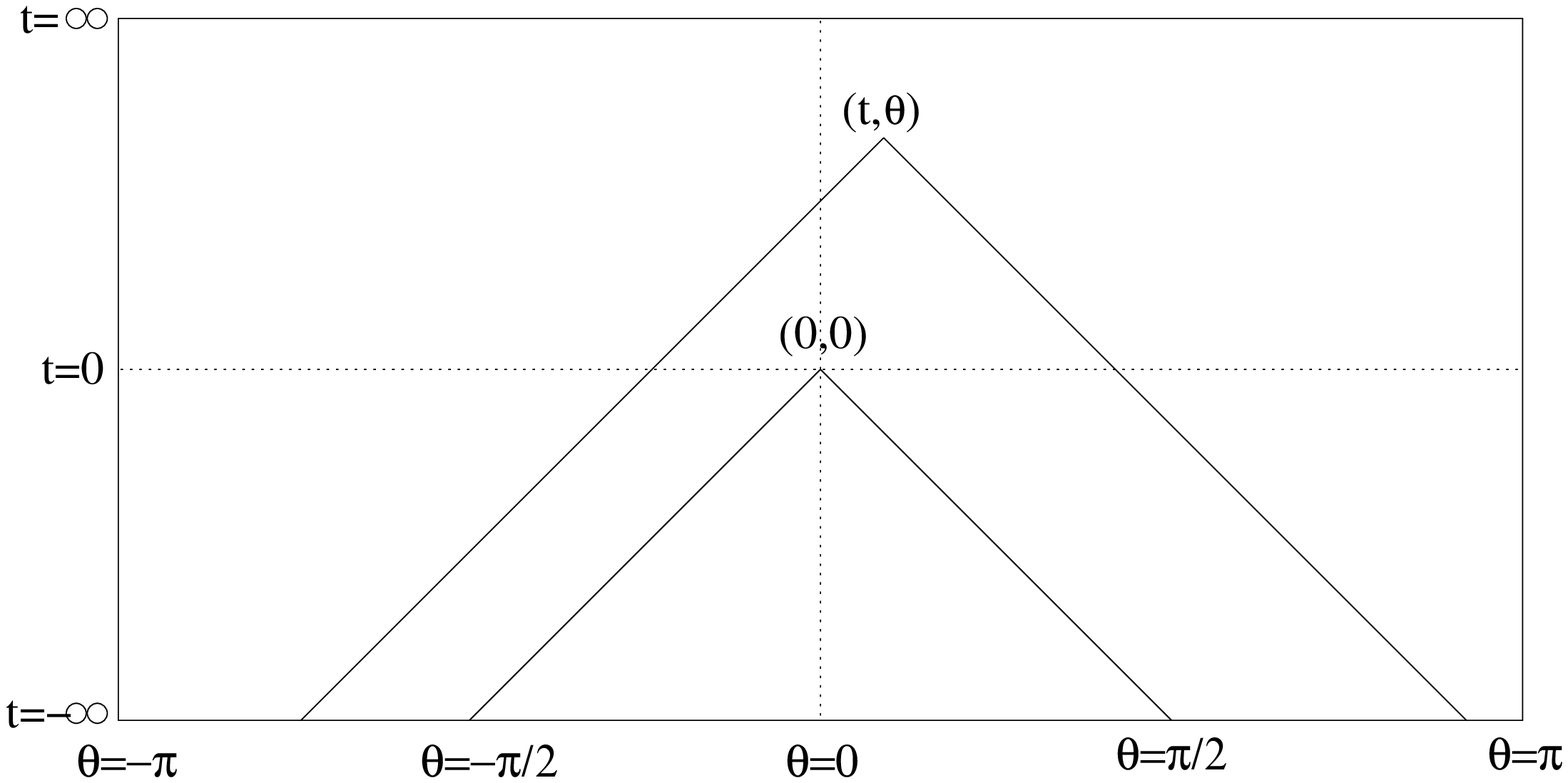,width=9cm}{\label{fig2}
	The Carter-Penrose diagram 
	of the two dimensional dS space. We have represented the causal past of the point $(t,\theta)$.}
	

As it will become clear in the sequel, there is another interesting way to express the general solution of
eq.(\ref{k2}). This takes into account the cuts at the bifurcating horizons $\phi=\pm {\pi \over 2}$.
Introducing the complex variable $z=\cos\phi$, the general solution can be written as
\be
   \Psi_0(z)=C\,(z-i\epsilon)^{i\mu-{1\over 2}}
   +D\,(z+i\epsilon)^{i\mu-{1\over 2}}\,,
\ee
where the limit
$\epsilon\rightarrow 0^+$ is understood. The function
$(z-i\epsilon)^{i\mu-{1\over 2}}$ is holomorphic in the lower half
plane. The constants are related by
\be
   A=C+D\,,\qquad
   B=(e^{-i\pi})^{i\mu-\frac12}\,C+(e^{i\pi})^{i\mu-\frac12}\,D\,.
   \label{ABCD}
\ee

To exploit the arbitrariness of the $A$ and $B$ coefficients, we
 introduce the following  operators:
\be
   a^{\dagger}(\phi)={1\over
   \sqrt{2\pi}}\sum_{m=-\infty}^\infty e^{im\phi}a^{\dagger}_m \,.
\ee
which create a position eigenstate when acting on the invariant vacuum: $a^\dagger(\phi)|\Omega\rangle=|\phi\rangle$.
The field at the origin can then be written as
\be
   \Phi(0,0)=A\int_{-\pi\over
   2}^{\pi\over 2} \ {d\phi}\, (\cos\phi)^{i\mu-{1\over
   2}}a^{\dagger}(\phi)+ B\int_{\pi\over 2}^{3\pi\over 2} \ {d\phi}\,
   (-\cos\phi)^{i\mu-{1\over 2}}a^{\dagger}(\phi)+{\rm h.c.}
   \label{zero}
\ee
Let us  define $\Phi_A$, the creation part of the field operator
with support inside the horizon, by
\be
   \Phi^+_A(0,0)=\int_{-\pi\over
   2}^{\pi\over 2}{d\phi}\, (\cos\phi)^{i\mu-{1\over
   2}}a^{\dagger}(\phi)\,.
\ee 
Then eq.(\ref{zero}) can be written as 
\be 
	\Phi(0,0)=A\,\Phi_A^+(0,0)+B\,e^{i\pi J}
	\Phi_A^+(0,0)e^{-i\pi J}+{\rm h.c.} 
\ee 
Therefore, by direct computation,
we get that the  field operator $\Phi(x)$ is given, at any point $x=(t,\theta)$,  by 
\be
   \Phi(x)=A\,\Phi_A^{+}(x)+B\,\Phi_A^{+}(\bar x) +{\rm h.c,}
   \label{ab}
\ee 
where $\bar x$ is the antipodal point of coordinates $(-t,\pi+\theta)$. Notice that we have
$\Phi_{A,B}(x)=\Phi_{B,A}(\bar x)$. 

In brief eq.(\ref{ab}) demonstrates that the general solution of the
covariance condition can be written as a superposition of two local fields living on either side of the
\emph{transported} horizon as explained in Figure \ref{fig2}. At this point, it is interesting to notice that when $A=B$,
 the field operator is well defined on the orbifold $dS_2/{\mathbb{Z}}_2$,
the elliptic de Sitter space \cite{Schr} which has recently received some attention \cite{EdS}.
As we will show in the next Section,
this choice is not compatible with the equal time commutation relations,
in other words, the corresponding field operator is not canonical.

We can now relate the Fourier coefficients $c_m$  to the above position representation.
The constants $c_0$ and $c_1$, which fully determine the solution,
can be directly related to the constants $A$ and $B$.
In fact, using the integral:
\be
   \int_{0}^{\pi/2}d\phi \, (\cos\phi)^{\gamma} ={\Gamma\left({1\over 2}\right)
   \Gamma\left({1+\gamma \over 2}\right) \over 2\,
   \Gamma\left({\gamma\over 2}+1\right)}\,,
\ee
 we have,
\be
   c_0={\gamma_0\over \sqrt{2}}\frac{\Gamma\left(i{\mu\over 2}+{1\over 4}\right)}
   {\Gamma\left(-i{\mu\over 2}+{1\over 4}\right)}(A+B),\qquad
   c_1=-i{\gamma_1\over \sqrt{2}}\frac{\Gamma\left(i{\mu\over 2}+{3\over 4}\right)}
   {\Gamma\left(-i{\mu\over 2}+{3\over 4}\right)}(A-B)\,.
   \label{cAB}
\ee

\subsection{Commutation relations}

In the preceding section, we showed that the covariance properties
determine the field operator up to two complex constants $c_0$ and
$c_1$,  or equivalently  $A$ and $B$. Here, we calculate the  equal time
commutators between two fields  and
between the field and its time derivative.
It turns out that the first commutator vanishes for all constants
and that the second is proportional to a delta function, as in the case of the Poincar\'e group.
Therefore any field, solution of eq.(\ref{e1dS}), is canonical. 
Similarly to what was done is Section 2, it is then possible to find the
identification of $\hbar$. The novelty is that the field is no longer unique.
We shall show indeed that the moduli space of solutions is non trivial since
it is given by SU$(1,1)/$U$(1)$.
In addition, the two commutators and the Klein-Gordon equation imply that
the canonical fields are causal. 

Consider first
\be
   g(t,\theta)=[\,\Phi(t,0),\,\Phi(t,\theta)\,]\,.
\ee
It is a c-number. Multiplying  this expression on the right by $e^{-i\theta J}$ and
on the left by $e^{i\theta J}$ we get
\ba
   g(t,\theta)&=&e^{-i\theta J}\,g(t,\theta)\,e^{i\theta J}
   \nonumber\\
   &=&[\,\Phi(t,-\theta),\,\Phi(t,0)\,]
   \nonumber\\
   &=&-g(t,-\theta)\,.
\ea
On the other hand, multiplying by the parity operator $\mathsf{P}$ we have
\ba
   g(t,\theta)&=&\mathsf{P}\,g(t,\theta)\,\mathsf{P}^{-1}
   \nonumber\\
   &=&[\,\Phi(t,0),\,\Phi(t,-\theta)\,]
   \nonumber\\
   &=&g(t,-\theta)\,.
\ea
From these two equations we get
\be
   [\,\Phi(t,0),\,\Phi(t,\theta)\,]=0\,.
\ee
The field operators at equal time thus
commute for all choices of the constants $c_0$
and $c_1$.

Consider next
\be
   h(\theta)=[\,\partial_t\Phi(0,0),\,\Phi(0,\theta)\,]\,.
\ee
We have the following equalities
\ba
   h(\theta)&=&i\,\Big[\,[\,K_1,\,\Phi(0,0)\,],\,\Phi(0,\theta)\,\Big]
   \nonumber\\
   &=&i\,\Big[\,e^{i\theta J}[\,K_1,\,\Phi(0,0)\,]e^{-i\theta J},\,\Phi(0,0)\,\Big]
   \nonumber\\
   &=&i\,\Big[\,[\,\cos\theta K_1-\sin\theta K_2,\,\Phi(0,-\theta)\,],\,\Phi(0,0)\,\Big]\,,
\ea
where we have used the definition of $\Phi(t,\theta)$ in terms
of $\Phi(0,0)$, the c-number character of $h(\theta)$ and the
transformation of $K_1$ under a rotation. Now use the Jacobi
identity as well as the previously proved identity $g(0,\theta)=0$
and $[K_2,\Phi(0,0)]=0$ to obtain
\ba
   h(\theta)&=&i\,\Big[\,[\,\cos\theta K_1,\,\Phi(0,0)\,],\,\Phi(0,-\theta)\,\Big]
   \nonumber\\
   &=&\cos\theta\, h(-\theta)\,.
   \label{ccr1}
\ea
 On the other hand, we have
\ba
   h(\theta)&=&i\,\Big[\,{\mathsf{P}}[\,K_1,\,\Phi(0,0)\,]
   {\mathsf{P}}^{-1},\,{\mathsf{P}} \Phi(0,\theta){\mathsf{P}}^{-1}\,\Big]
   \nonumber\\
   &=&i\,\Big[\,[\,K_1,\,\Phi(0,0)\,],\,\Phi(0,-\theta)\,\Big]
   \nonumber\\
   &=&h(-\theta)\,,
   \label{cp}
\ea
where we used ${\mathsf{P}}K_1{\mathsf{P}}^{-1}=K_1$. Finally, using eq.(\ref{ccr1}) and eq.(\ref{cp}), we
obtain 
\be
    (1-\cos\theta)\,h(\theta)=0\,,
\ee
with $h(\theta)$ even in $\theta$.
Since we work on the circle, the solution is
\be
   h(\theta)=iN\,\delta(\theta)\,.
\ee
The constant  $N$ is real since $\Phi(t,\theta)$ is hermitian.
Therefore, unless $N$ is equal to zero the covariant field $\Phi$ is  canonical.

We now determine $N$ in terms of $c_0$ and $c_1$. We need
\be
   [\,K_1,\,a_m^\dagger\,]={i\over 2}\left[\left( m -i\mu+{1\over 2}\right)a^\dagger_{m+1}
   -\left( m +i\mu-{1\over 2}\right)a^\dagger_{m-1}\right],
\ee
in order to calculate $[\,K_1,\,\Phi(0,0)\,]$.
Since
\be
   \int_{0}^{2\pi}{d\theta\over 2\pi}\,\Phi(0,\theta)=c_0\,a^{\dagger}_{0}+c_0^*\,a_0\,,
\ee
and $c_{-1}=c_1$ we get
\be
   {N\over 2\pi}=-2\,{\rm Re}\left[\left(\mu-{i\over 2}\right)c_0^*c_1\right].
\ee

The Planck constant $\hbar$ is introduced as $N=-\hbar$. In the
following it will be set to one and thus
\be
   {\rm Re}\left[\left(\mu-{i\over2}\right)c_0^*c_1\right]={1\over 4\pi}\,.
   \label{ccr}
\ee
Defining the 2-component vector:
\be
   {\mathbf{z}}={c_0\over \gamma}\,\sqrt{2\pi}{1\choose1}
   \,+\,\gamma^*c_1\left(\mu-\frac i2\right)\,\sqrt{2\pi}{1\choose-1}\,,
   \label{defz}
\ee 
where $\gamma$ is an arbitrary complex constant, eq.(\ref{ccr}) reads 
\be
   {\mathbf{z^{\dagger}\sigma_3 z}}=1\,,
   \label{ccrz}
\ee 
where ${\mathbf{\sigma_3}}=diag(1,-1)$ is the third Pauli matrix. The invariance group of this equation is
U$(1,1)$. Since a fixed two component vector is invariant under a subgroup U$(1)$, the solutions of
eq.(\ref{ccr}) parameterize the coset U$(1,1)/$U$(1)$. In addition, the overall phase in $\bf z$ has no physical
meaning, the moduli space is thus SU$(1,1)/$U$(1)$. We notice that this is also the moduli space of the
alpha vacua \cite{Allen:1985ux}. This connection will be made more precise in the next section. 
A convenient parameterization of $\bf z$ is given by 
\be
   {\mathbf{z}}=\binom{\cosh\alpha}{e^{i\beta}\sinh\alpha}\,,
   \label{paraz}
\ee
where the two components will be shown to be Bogoliubov coefficients when interpreted
in the usual field theoretical description.

\subsection{Time reversal}

One gains a better understanding of the field properties when considering the discrete transformations which leave the point $(0,0)$ invaraint.

As we have already seen, invariance under parity $\phi\rightarrow -\phi$
is automatically realized since  $c_{-m}=c_m$ for all choices of $c_0$ and $c_1$.

Invariance under time reversal is more instructive. It is implimented by
${\mathsf{T}} e^{i\theta J}{\mathsf{T}}^{-1}=e^{i\theta J}$
and ${\mathsf{T}} e^{i\omega^aK_a}{\mathsf{T}}^{-1}=e^{-i\omega^aK_a}$.
It can be realized in two different ways \cite{rep4}.
The first one is by a unitary operator as
\be
   {\mathsf{T}}|m\rangle=\epsilon(-1)^m|m\rangle\,,
\ee
with $\epsilon=\pm 1$. The unitary time inversion operator acts on the field as
\be
   {\mathsf{T}}\Phi(0,0){\mathsf{T}}^{-1}=\sum_{m=-\infty}^{\infty}
   (-1)^m c_m\,a^\dagger_m + (-1)^m c_m^*\,a_m\,,
\ee
that is  ${\mathsf{T}}\Phi(t,\theta){\mathsf{T}}^{-1}=\Phi(-t,\theta+\pi)$.
If we demand that ${\mathsf{T}}\Phi(0,0){\mathsf{T}}^{-1}=\Phi(0,0)$ then we get
that $c_{2m+1}=0$  for $\epsilon=1$ or $c_{2m}=0$ for $\epsilon=-1$.
These two relations  lead to a vanishing $N$  and so are incompatible with the requirement of the
canonical commutation
relations, so we must use the second way of realizing the time
reversal  invariance.

The second one is, as in flat space-time,
by an anti-unitary operator. In this case,  ${\mathsf{T}}$ acts on the basis $|m\rangle$ as
\be
   {\mathsf{T}}|m\rangle=e^{i\nu_m}|-m\rangle\,,
   \label{TR}
\ee
with
\be
   e^{i(\nu_{m+1}-\nu_m)}=-{m+{1\over 2}-i\mu\over
   m+{1\over 2}+i\mu}={\gamma_{m+1}\over \gamma_{m+1}^*}
   {\gamma_{m}^*\over \gamma_{m}}\,.
\ee
The last relation implies that
\be
   e^{i\nu_m}=e^{i\delta}{\gamma_m\over \gamma_m^*}\,,
\ee
where $\delta$ is common unobservable phase which can be put to zero without loss of generality.
When acting on the field we get
\be
   {\mathsf{T}}\Phi(0,0){\mathsf{T}}^{-1}=\sum_{m=-\infty}^{\infty}
   c_m^*e^{i\nu_m}\,a^\dagger_m + c_me^{-i\nu_m}\,a_m\,,
\ee
That is the field ${\mathsf{T}}\Phi(0,\theta){\mathsf{T}}^{-1}$ is of the same form
as $\Phi(0,\theta)$ with the coefficients $c_m$ replaced by $c_m^*e^{i\nu_m}$:
\be
   {\mathsf{T}}\Phi(t,\theta)_{\{c_m\}}{\mathsf{T}}^{-1}
   =\Phi(-t,\theta)_{\{c_m^*e^{i\nu_m}\}}\,.
\ee
Notice that $c_m^*e^{i\nu_m}$ verify the defining relations of the
$c_m$ as they should. 

If we require ${\mathsf{T}}\Phi(0,0){\mathsf{T}}^{-1}=\Phi(0,0)$
then we get
\be
   c^*_me^{i\nu_{m}}=c_{m}\,.
   \label{tr}
\ee 
This implies that the phase of $c_m$ is $\nu_m/2$. This condition is compatible with eq.(\ref{inva}), and
also fixes the  physically meaningful relative phase between $c_0$ and $c_1$. 
Explicitely, the anti-unitary time inversion invariance gives
\be
   {c_1c_0^*\over{c_1^*c_0}}=-{{1\over 2}-i\mu\over
   {1\over 2}+i\mu}\,.
\ee
This condition can be written as
\be
   {\rm Im}\left[\left(\mu-{i\over 2}\right)c_0^*c_1\right]=0\,.
   \label{trvs}
\ee

The canonical commutation relation and time reversal can thus be regrouped in a single simple
formula:
\be
   \left(\mu-{i\over 2}\right)c_0^*c_1=\frac1{4\pi}\,.
\ee 
In terms of the parameterization of eq.(\ref{paraz}), we simply get $\beta=0$.

\section{Two point function and Bunch-Davies vacuum}

As briefly explained in the Introduction, the easiest way to make
contact with the usual treatment of quantum fields on dS space is through
computation of the two point-function which is an observable.

In the present approach, the two point function evaluated in the unique
invariant vacuum state is given by
\be
   \langle\Omega|\Phi(0,0)\Phi(0,\theta)|\Omega\rangle=
   \sum_{m=-\infty}^\infty|c_{m}|^2e^{-im\theta}\,.
\ee
Imposing the Hadamard form of this two point function,
that is, the same behavior in the coincidence point limit
 as that of the flat case, gives a constraint
on the asymptotic value of the $c_m$.
Recall that the flat two point function on the circle is given by
\be
   {1\over 4\pi}\sum_{m=-\infty}^\infty{1\over \sqrt{m^2+M^2}}e^{-im\theta}\,.
\ee
At short distances, it reduces to
\be
   -{1\over 2\pi}\log\theta\,.
\ee 
The asymptotic behavior of $c_m$ can be determined from eq.(\ref{gam}). One finds
\be
   |c_{2m}|^2= \frac{|c_0|^2}{|\gamma_0|^2}\,{|\gamma_{2m}|^2}
   \underset{m\to\infty}{\approx}{|c_0|^2\over|\gamma_0|^2}\,\frac1{|m|}\,
   ,\qquad
   |c_{2m+1}|^2= \frac{|c_1|^2}{|\gamma_1|^2}\,{|\gamma_{2m+1}|^2}
   \underset{m\to\infty}{\approx}{|c_1|^2\over|\gamma_1|^2}\,\frac1{|m|}\,.
\ee
This has the dependence on $m$ required so that  the two point function coincides with
 the flat one at short distances. Indeed we get 
\be
   -2\left(\left|{c_0\over \gamma_0}\right|^2+\left|{c_1\over \gamma_1}\right|^2\right)\log\theta\,.
\ee
Comparison of the overall constants gives
\be
   \left|{c_0\over \gamma_0}\right|^2+\left|{c_1\over \gamma_1}\right|^2={1\over 4\pi}\,.
\ee 
From this equation and   the canonical commutation relation which reads
\be
	{\rm Re}\left[\left(\frac{c_0}{\gamma_0}\right)^*\left(\frac{c_1}{\gamma_1}\right)\right]={1\over 8\pi}\,,
\ee
we get 
\be
	\left|{c_0\over \gamma_0}-{c_1\over \gamma_1}\right|^2=0\,.
\ee
This gives (up to a choice of the unphysical phase) 
\be
   c_0^{\mathrm{BD}}={1\over \sqrt{8\pi}}\,\gamma_0\,,\qquad
   c_1^{\mathrm{BD}}={1\over \sqrt{8\pi}}\,\gamma_1\,.
\ee 
Therefore the field operator is characterized for all $m$ by 
\be
   c_m^{\mathrm{BD}}={1\over \sqrt{8\pi}}\,\gamma_m\,.
   \label{cBDs}
\ee
We have called this solution Bunch-Davies (BD) because it 
possesses a fixed sign of 
the conformal frequency in the large $m$ limit, see next section for further discussion.
In the limit of large $\mu$ we obtain
\be
   c_0^{\mathrm{BD}},\,c_1^{\mathrm{BD}}\underset{\mu\to\infty}{\approx}
   {1\over {\sqrt{4\pi\mu}}}\,,
\ee
which is expected from  flat space. Notice that the BD vacuum is time reversal  invariant, 
in the sense of eq.(\ref{tr}).

In terms of ($A,B$) and ($C,D$), the BD field operator is characterized, up to an overall phase, by
 \be
       \begin{array}{ll}
   	A^{\mathrm{BD}}=
   	\frac{e^\frac{\pi\mu}2}{\sqrt{8\pi\cosh\pi\mu}}\,,
   	&B^{\mathrm{BD}}=\frac{-ie^{-\frac{\pi\mu}2}}{\sqrt{8\pi\cosh\pi\mu}}
   	=-ie^{-\pi\mu}A^{\mathrm{BD}}\,,\\ &\\
   	C^{\mathrm{BD}}=0\,,
   	&D^{\mathrm{BD}}=\frac{e^\frac{\pi\mu}2}{\sqrt{8\pi\cosh\pi\mu}}\,.\\
       \end{array}
\ee
From the last two equations, we learn that $D$ ($C$) is the amplitude of positive (negative) conformal
frequency modes.

Using the BD solution as the reference solution in eq.(\ref{defz}), which amounts to choose $\gamma=\gamma_0$,
the moduli space of canonical fields, eq.(\ref{paraz}), 
can be parameterized by 
\be
\begin{array}{ll}
   c_0=c_0^{\mathrm{BD}}(\cosh \alpha+e^{i\beta}\sinh \alpha)\,,
   &c_1= c_1^{\mathrm{BD}}(\cosh \alpha-e^{i\beta}\sinh \alpha)\,,\\
   &\\
   A=\cosh \alpha \,A^{\mathrm{BD}}+
   e^{i\beta} \sinh \alpha \,B^{\mathrm{BD}}\,,
   &B=\cosh\alpha  \,B^{\mathrm{BD}}+e^{i\beta} \sinh \alpha\, A^{\mathrm{BD}}\,,\\
   &\\
   C=-i\,e^{i\beta}\sinh\alpha \,e^{-\pi\mu}\, D^{\rm BD}\,,
   &D=\cosh\alpha \,D^{\rm BD}.\\
   \end{array}
   \label{cBD}
\ee
Using eq.(\ref{ab}) and defining the operator:
\be
   \Phi_{\mathrm{BD}}^+(x)=A^{\mathrm{BD}}\,\left(\,\Phi_A^+(x)-ie^{-\pi\mu}\Phi_A^+(\bar x)\,\right)\,,
\ee
the general field can be expressed as
\be
   \Phi_{\alpha,\beta}(x)=\cosh\alpha\,\Phi_{\mathrm{BD}}^+(x)+e^{i\beta}\sinh\alpha
   \,\Phi_{\mathrm{BD}}^+(\bar x)+{\rm h.c.}
\label{gener}
\ee

Let us denote the BD positive Wightman function
by
\be
	G_0(x;y)=\langle\Omega\vert\,\Phi_{\rm{BD}}(x)\,
	\Phi_{\rm{BD}}(y)\,\vert\Omega\rangle\,.
\ee
It satisfies 
$G_0(y;x)=G_0^*(x;y)$ and $G_0(\bar x;\bar y)=G_0(y;x)$. 
Using eq.(\ref{gener}), 
we can express the positive Wightman function of any field operator
in terms of $G_0(x;y)$:
\ba
   	G_{\alpha,\beta}(x;y)&=&\langle\Omega\vert\,\Phi_{\alpha,\beta}(x)\,
	\Phi_{\alpha,\beta}(y)\,\vert\Omega\rangle
	\nonumber\\
	&=&\cosh^2\alpha\, G_0(x;y)\,+\,\sinh^2\alpha\, G_0(\bar{x};\bar{y})
	\nonumber\\
	&&+\,\sinh\alpha\cosh\alpha\left( e^{i\beta}\, G_0(x;\bar y)
	\,+\,e^{-i\beta}\, G_0(\bar x;y)\right)\,.
   	\label{pwtf}
\ea
This expression is identical to eq.(2.15) in \cite{Allen:1985ux} which gives the relation between the Wightman function evaluated in 
the alpha vacuum characterized by $\alpha$ and $\beta$, and that evaluated in 
the Bunch-Davies vacuum. The 
relationship between our approach and QFT in dS space will be further analyzed in section 7.

To conclude we notice that, in the present formalism,
 we can also consider the Wightman function between two different field operators,
namely
\ba
   \langle\Omega\vert\,\Phi_{\alpha,\beta}(x)\,\Phi_{\alpha',\beta'}(y)\,\vert\Omega\rangle &=&
   \cosh\alpha\cosh\alpha' \,G_0(x;y)\,+\,e^{-i(\beta-\beta')}
   \sinh\alpha\sinh\alpha' \,G_0(\bar{x};\bar{y})
   \nonumber\\
   &&+\,e^{i\beta'}\cosh\alpha\sinh\alpha' \,G_0( x;\bar y)
   \,+\,e^{-i\beta}\cosh\alpha'\sinh\alpha \,G_0(\bar x;y)\,.\qquad\quad
  \label{gener2}
\ea 
It is then a legitimate question
 to ask in which circumstances this type of expression
could occur in QFT. To get it, one should consider 
the cross term of the response function matrix
associated with two local quantum systems
coupled to two different fields.
More explicitely, one should consider interaction Hamiltonians 
of the form (in the interacting picture):
\be
	H_{int}= g_1\, q_1(t) \,  \Phi_{\alpha,\beta}(t,{\bf x}) + 
	g_2\, q_2(t) \,  \Phi_{\alpha',\beta'}(t,{\bf y})\,, 
\ee
where $q_1, q_2$ are coordinates of the two systems and where $g_1$
and $g_2$ are the coupling constants. Then, to first order
in $g_1 g_2$, the amplitude to induce transitions to both oscillators 
will be governed by eq.(\ref{gener2}).

\section{{\it in} and {\it out} vacuum}

The explicit transformation of the state vector in the UIR
recalled in Section 2 can be used to determine the full space-time
dependence of the field operator, and this without using the Klein-Gordon equation.
One can then use the asymptotic behavior of the field
operator to identify the {\it in} and {\it out} positive frequency solutions.

In order to determine $\Phi(t,0)$
let us define  the vector state $|\Psi_t\rangle=\Phi(t,0)|\Omega\rangle$.
It is the boosted state $e^{it K_1}|\Psi_0\rangle$.
In the $\phi$-representation, using eq.(\ref{te}), we get that $\Psi_t(\phi)=\langle\phi|\Psi_t\rangle$ is
\ba
   	\Psi_t(\phi)&=&A\,\Theta(\cos\phi+\tanh t)\,(\cos\phi \cosh t+\sinh t)^{i\mu-{1\over 2}}
   	\nonumber\\
	&&\,+\,B\,\Theta(-\cos\phi-\tanh t)\,(-\cos\phi\cosh t-\sinh
	t)^{i\mu-{1\over 2}}\,.\label{evoo}
\ea
The support of the term proportionnal to $A$ is
the space region in the causal past of $(t, 0)$ 
(see Figure \ref{fig2}).
From eq.
(\ref{evoo}), the field $\Phi(t,0)$ is 
\ba
   \Phi(t,0)&=&\int_0^{2\pi} d\phi\,\Psi_t(\phi)\,a^\dagger(\phi)+\Psi_t^*(\phi)\,a(\phi)
   \nonumber\\
   &=&A\,\int_{-\pi+\arccos(\tanh t)}^{\pi-\arccos(\tanh t)}{d\phi}\,
   {(\cosh t \cos \phi+\sinh t )^{i\mu-{1\over 2}}}
   \ a^{\dagger}(\phi)+{\rm h.c.}
   \nonumber\\
   &&+\,B\,\int_{\pi-\arccos(\tanh t)}^{\pi+\arccos(\tanh t)}{d\phi}\,
   {(-\cosh t \cos \phi-\sinh t )^{i\mu-{1\over 2}}}\ a^{\dagger}(\phi)+{\rm h.c.}
\ea
Performing a rotation of angle $\theta$ we obtain
\ba
   \Phi(t,\theta) &=&A\int_{-\pi+\arccos(\tanh t)}^{\pi-\arccos(\tanh t)}{d\phi}\,
   {(\cosh t \cos\phi+\sinh t )^{i\mu-{1\over 2}}}
   \ a^{\dagger}(\phi+\theta)+{\rm h.c.}
   \nonumber\\
   &&+\,B\int_{\pi-\arccos(\tanh t)}^{\pi+\arccos(\tanh t)}{d\phi}\,
   {(-\cosh t \cos\phi-\sinh t )^{i\mu-{1\over 2}}}
   \ a^{\dagger}(\phi+\theta)+{\rm h.c.}
\ea

To determine the field operators
propagating positive frequencies in the infinite proper past and future
it will be convenient to decompose the field in  Fourier modes:
\be
   \Phi(t,\theta)=\sum_{m=-\infty}^\infty c_m(t)\ e^{-im\theta }
   a^{\dagger}_m+c_m^*(t)\ e^{im\theta }a_m\,.
\ee
We get
\ba
   c_m(t)&=&A\int_{-\pi+\arccos(\tanh t)}^{\pi-\arccos(\tanh t)}
   {d\phi\over{\sqrt{2\pi}}}
   {(\cosh t \cos\phi+\sinh t )^{i\mu-{1\over 2}}}
   e^{-im\phi} 
   \nonumber\\
   &&+\,B\int_{\pi-\arccos(\tanh t)}^{\pi+\arccos(\tanh t)}
   {d\phi\over{\sqrt{2\pi}}}
   {(-\cosh t \cos\phi-\sinh t )^{i\mu-{1\over 2}}}
   e^{-im\phi}\,.
   \label{fm}
\ea

In terms of the conformal time
$\eta$ defined by $e^t=\tan\frac\eta2$, we have
\ba
   c_m(\eta)&=&{A\ (\sin \eta)^{-i\mu+{1\over 2}}}
   \int_{-\eta}^{\eta} {d\phi\over{\sqrt{2\pi}}}
   (\cos\phi-\cos\eta)^{i\mu-{1\over 2}}e^{-im\phi}
   \nonumber\\
   &&+\,B\ (\sin \eta)^{-i\mu+{1\over 2}}\int_{\eta}^{2\pi-\eta}
   {d\phi\over{\sqrt{2\pi}}}
   (-\cos\phi+\cos\eta)^{i\mu-{1\over 2}}e^{-im\phi}\,.
\ea
The integrals are ``Mehler-Dirichlet" integral representations of the
associated Legendre functions \cite{bateman}. Explicitely, one has
\be
   c_m(\eta)={  (\sin \eta)^{{1\over 2}}}\,\Gamma\left(i\mu+{1\over 2}\right)
   \left(
   A\,{\rm P}^{-i\mu}_{-m-{1\over 2}}(\cos \eta)
   +(-1)^m B\,
   {\rm P}^{-i\mu}_{-m-{1\over 2}}(-\cos \eta)\right).
\ee

The limits $t\to \pm \infty$ correspond to
$z=\cos\eta\to \mp 1$. In these limits, the asymptotic behavior
of the associated Legendre function is given by
(see page 164 in \cite{bateman})
\ba
   {\rm P}^{-i\mu}_{m-{1\over 2}}(z)&\underset{z\to 1}{\approx}&{2^{-\frac{i\mu}2}\over\Gamma(1+i\mu)}
   (1-z)^{\frac{i\mu}2}\,,
   \nonumber\\
   {\rm P}^{-i\mu}_{m-{1\over 2}}(z)&\underset{z\to -1}{\approx}&{2^{\frac{i\mu}2}\Gamma(i\mu)\over
   \Gamma\left(m+{1\over 2}+i\mu\right)\Gamma\left(-m+{1\over 2}+i\mu\right)}
   (1+z)^{-\frac{i\mu}2}+(-1)^m {2^{-\frac{i\mu}2}\Gamma(-i\mu)\over \pi}(1+z)^{{\frac{i\mu}2}}\,.
   \nonumber\\
\ea
The asymptotic behavior of $c_m(t)$ in the early past
can thus be determined as
\ba
   &&c_m(t)\underset{t\to-\infty}{\approx}\sqrt2\,e^{t\over 2}\,\Gamma\left(i\mu+{1\over 2}\right)\times
   \nonumber\\
   &&\,\times\left[\,A\,\frac1{\Gamma(1+i\mu)}e^{i\mu t}+
   B\left({(-1)^m \Gamma(i\mu)\over\Gamma\left(m+{1\over 2}+i\mu\right)
   \Gamma\left(-m+{1\over 2}+i\mu\right)}e^{-i\mu t}
   +\frac{\Gamma(-i\mu)}\pi e^{i\mu t}\right)\,\right].\nonumber\\
\ea
The solution with positive proper time
frequency in the remote past thus corresponds to
\be
   B^{\rm{IN}}=0\,.
\ee
This characterizes the {\it in} vacuum \cite{Mottola:1984ar}.

The asymptotic behavior of $c_m(t)$
 in the late future can similarly be deduced from
\ba
   &&c_m(t)\underset{t\to\infty}{\approx}\sqrt2\,e^{-{t\over 2}}\,\Gamma\left(i\mu+{1\over 2}\right)\times
   \nonumber\\
   &&\,\times\left[\,A\,\left({\Gamma(i\mu)\over
   \Gamma\left(m+{1\over 2}+i\mu\right)\Gamma\left(-m+{1\over 2}+i\mu\right)}e^{i\mu t}
   +{(-1)^m \over \pi}\Gamma(-i\mu)e^{-i\mu t}\right)+B\,\frac{(-1)^m}{\Gamma(1+i\mu)}
   e^{-i\mu t}\right].
   \nonumber\\
\ea
The {\it out} vacuum is thus characterized by
\be
   A^{\rm{OUT}}=\frac{-\pi\,B^{\rm{OUT}}}
   {\Gamma(-i\mu)\Gamma(1+i\mu)}=i\sinh\pi\mu\,B^{\rm{OUT}}\,.
\ee
It is easy to obtain the {\it in} to {\it out} transformation when noticing 
that {\it in} and {\it out} vacua are time-reversal states, which implies
$\alpha^{\rm IN}=\alpha^{\rm OUT}$ and $\beta^{\rm IN}=-\beta^{\rm OUT}$,
or ${\bf z}^{\rm OUT}={\bf z}^{\rm IN*}$. Therefore, 
using the fact that $\beta^{\rm IN} = {\pi \over 2} $ and $\tanh \alpha^{\rm IN} =
e^{-\pi\mu}$,  the mean number of {\it out}
quanta of momentum $m$ in {\it in} vacuum is
\ba
	\bar n_{\rm OUT}&=&\vert\, e^{i\beta^{\rm IN}}\sinh\alpha^{\rm IN}
	\cosh\alpha^{\rm IN}+{\rm c.c.}\,\vert^2
	\nonumber\\
	&=&\frac1{\sinh^2\pi\mu}\,,
\ea
which coincide with the 4-dimensional result, see eq.(32) in \cite{Mottola:1984ar}.

The  large $m$ behavior of $c_m(\eta)$ can also be determined as
\be
   c_m(\eta)\underset{m\to\infty}{\approx}\frac{e^{-i{\pi\over 4}}e^{\mu\pi\over 2}}{\sqrt{2\pi\,}m!}
   {\Gamma\left(m+{1\over 2}-i\mu\right)\Gamma\left(i\mu+{1\over 2}\right)}
   \left[e^{im\eta}(A+ie^{-\mu\pi}B)+e^{-im\eta}(B+ie^{-\mu\pi}A)\right].
\ee
This clearly exhibits the defining property of the Bunch-Davies vacuum
as having only positive conformal frequency for large momenta.
Notice that this equation can also be expressed in terms of the coefficients $C$
and $D$ as
\be
   c_m(\eta)\underset{m\to\infty}{\approx}\frac{e^{-i{\pi\over 4}}e^{\mu\pi\over 2}}{\sqrt{2\pi}\,m!}
   {\Gamma\left(m+{1\over 2}-i\mu\right)\Gamma\left(i\mu+{1\over 2}\right)}
   \left[e^{im\eta}(1+e^{-2\pi \mu})\,D+2ie^{-im\eta}\cosh \pi\mu\, C\right].
\ee
This shows the holomorphic properties of the Bunch-Davies vacuum: $C=0$.

\section{Relation to field theory approach}

The aim of this section is the following. Starting with the usual treatment of QFT on dS space, 
we construct a squeezing operator which transforms  an operator annihilating an alpha vacuum into an
operator annihilating another vacuum.
When acting on the first alpha vacuum this operator engenders the second vacuum.
We shall show that when it acts on the field operator corresponding to the first vacuum, 
one obtains the  canonical field associated with the second one.

In field theory, the free scalar field is first decomposed
into Fourier modes
\be
    \Phi(t,\theta)={1\over \sqrt{2\pi}}\sum_{m=-\infty}^\infty
    a_m\,u_m(t)\,e^{im\theta}\,+\,
    a^{\dagger}_m\,u_m^*(t)\,e^{-im\theta}\,,
\ee
where the time dependent modes satisfy the
differential equation
\be
    \left(\,\frac{d^2}{dt^2}+\tanh t\frac{d}{dt}
    +\frac{m^2}{\cosh ^2t}\,\right)
    u_m(t)=-\left(\mu^2+\frac14\right)\, u_m(t)\,,
    \label{kkg}
\ee
and the Wronskian condition
\be
    \cosh t\Big(\,u^*_m(t)\,\dot{u}_m(t)\,-\,\dot{u}^*_m(t)\,u_m(t)\,\Big)=-i\,,
    \label{wr}
\ee
so that the equal time canonical commutation relations are satisfied.
When the boundary condition fixing $u_m$ is parity invariant, 
the situation we shall consider, one also has
\be
	u_m(t)=u_{-m}(t)\,.
\ee

The solution $u_m$ is not uniquely defined by the two requirements (\ref{kkg})
and (\ref{wr}).
In fact
\be
    \tilde u_m\,=\,e^{i\sigma_m}\cosh\rho_m\,u_m\,+\,e^{-i\lambda_m}\sinh\rho_m\,u_{-m}^*\,,
    \label{bgt}
\ee
also satisfies the two requirement, where $\rho_m$, $\sigma_m$ and $\lambda_m$ are arbitrary real constants.
Fixing the unphysical phase of  $\tilde u_m$, we take herefrom $\sigma_m$ to zero.
The corresponding Bogoliubov transformation is given by
\be
    \tilde a_m\,=\,\cosh\rho_m\,a_m\,-\,e^{i\lambda_m}\sinh\rho_m\, a^{\dagger}_{-m}\,.
\ee
This transformation can be obtained by introducing the so-called
squeezing operators
\be
	{\cal S}_{\rho_m,\lambda_m}=\left\{
    	\begin{array}{ll}
    		\exp\left[\,\frac{\rho_0}2\left\{\,e^{i\lambda_0}\,(a^{\dagger}_0)^2
    		\,-\,e^{-i\lambda_0}\,(a_0)^2\,\right\}\,\right],&{\rm for}\quad m=0\,,\\
		&\\
 		\exp\left[\,\rho_m\left\{\,e^{i\lambda_m}\,a^{\dagger}_m a^{\dagger}_{-m}
    		\,-\,e^{-i\lambda_m}\,a_m a_{-m}\,\right\}\,\right],&{\rm for}\quad m \neq 0\,.\\
	\end{array}
	\right.
\ee
 We then have (see e.g. \cite{mandel})
 \be
    \tilde a_m\,=\,{\cal S}_{\rho_m,\lambda_m}
    a_m\,{\cal S}_{\rho_m,\lambda_m}^{-1}\,.
 \ee
 
 The relation between the two vacua is formally given by
 the product over all $m$:
 \be
    |{\Omega}_{\{\rho_m,\lambda_m\}}\rangle\,=
    \,\left(\prod_{m=0}^{\infty}
    {\cal S}_{\rho_m,\lambda_m}\right)
    \, |\Omega\rangle\,.
    \label{vac}
 \ee
 The reason we wrote \emph{formally} is the following.
 The squeezing operator ${\cal S}_{\rho_m,\lambda_m}$ possesses the normal ordered form:
 \be
	{\cal S}_{\rho_m,\lambda_m}=
	\exp\left[\,\,e^{i\lambda_m}\tanh\rho_m\, a_m^\dagger a_{-m}^\dagger\,\right]\,
    	\left({1\over{\cosh\rho_m}}\right)^{1+a^{\dagger}_m a_m+a^\dagger_{-m} a_{-m}}\,
    	\exp\left[\,-\,e^{-i\lambda_m}\tanh\rho_m\, a_m a_{-m}\,\right].
\ee
The product over $m$ of the  squeezing operators contains an overall factor given by
\be
    \prod_{m=1}^\infty \frac1{\cosh\rho_m}\,.
    \label{sqc}
\ee 
 The product 
 is a well defined unitary operator provided the above coefficient is nonvanishing, 
 which implies $\rho_m^2\rightarrow0$ faster than $1/m$.

In general the vacuum $|{\Omega}_{\{\rho_m,\lambda_m\}}\rangle$
is not dS invariant, that is, the Wightman function in this vacuum:
\ba
    G_{\{\rho_m,\lambda_m\}}(t,\theta;t',\theta')&=&
    \langle\Omega_{\{\rho_m,\lambda_m\}}\vert\,\Phi(t,\theta)\,
    \Phi(t',\theta')\,\vert\Omega_{\{\rho_m,\lambda_m\}}\rangle
    \nonumber\\
    &=&\frac1{2\pi}\sum_{m=-\infty}^\infty
    \tilde{u}_m(t)\,\tilde{u}^*_m(t')\,e^{im(\theta-\theta')}
    \label{ftp}
\ea
is not a function of dS invariant quantities.
The requirement that the vacuum be dS invariant puts severe constraints on the
$\tilde{u}_m$. Since all the dS invariant vacua are related to each others by
Bogoliubov transformation (\ref{bgt}), this requirement
constrains $\rho_m$ to be $m$ independent and $\lambda_m$ to possess a given
$m$ dependence associated with the phase of the modes $u_m$.

If we choose the reference vacuum to be the Bunch-Davies vacuum, the 
time dependence at fixed $m$ is
\ba
    u_m(t)u^*_m(t')&=&\cosh^2\rho_m\,u^{\rm BD}_m(t)u^{\rm BD*}_{-m}(t')
    \,+\,\sinh^2\rho_m u^{\rm BD*}_m(t)u^{\rm BD}_{-m}(t')
    \nonumber\\
    &&+\,\sinh\rho_m\cosh\rho_m\Big(e^{i\lambda_m}\,u^{\rm BD}_m(t)u^{\rm BD}_{-m}(t')
    \,+\,e^{-i\lambda_m}u^{\rm BD*}_m(t)u^{\rm BD*}_{-m}(t')\Big)\,.\qquad
\ea
Then, if as in \cite{Allen:1985ux}, we choose the arbitrary phase by imposing
\be
    u^{\rm BD*}_{m}(t)=(-1)^m \,  u^{\rm BD}_{m}(-t)\,,
\ee
taking $\rho_m=\alpha$ and $\lambda_m=\beta$, the two point function (\ref{ftp}) evaluated in this
$(\alpha,\beta)$-vacuum coincides 
with the expression in  eq.(\ref{pwtf}).

In the formalism adopted in this paper, dS invariance is built in, and the field operators are 
decomposed
in Fourier modes given  in eq.(\ref{fm}). Using the time reversal invariance of the BD field operator as well
as the explicit expression of the time reversal operators given in eq.(\ref{TR}), we obtain
\be
    u^{\rm BD}_{m}(-t)=\frac{\gamma_m^*}{\gamma_m}\,u^{\rm BD*}_{m}(t)\,.
\ee
When taking $\rho_m=\alpha$ and $e^{i\lambda_m}=e^{i\beta}(-1)^m\frac{\gamma_m}{\gamma_m^*}$,
the two-point function (\ref{ftp}) gives again (\ref{pwtf}).
However, because of dS invariance, 
the normalization coefficient of eq.(\ref{sqc}) vanishes.  Different vacua are therefore
only formally related by the operator
${\cal S}_{\alpha,\beta}=\prod_{m=0}^\infty{\cal S}_{\{\alpha,\beta\}}$.
In fact, the different vacua belong to inequivalent Fock spaces.

Proceeding formally, the action of this operator 
${\cal S}_{\alpha,\beta}$ on the field $\Phi_{\rm BD}$ gives
\ba
    {\cal S}_{\alpha,\beta}^{\dagger}\,\Phi_{\rm BD}(0)\,{\cal S}_{\alpha,\beta}&=&
    \sum_{m=-\infty}^\infty c_m^{\rm BD}\,
     {\mathcal{S}}^{\dagger}_{\alpha,\beta}\,a^{\dagger}_m\,{\mathcal{S}}_{\alpha,\beta}
    \,+\,c_m^{\rm BD*}\,
     {\mathcal{S}}^{\dagger}_{\alpha,\beta}\,a_m\,{\mathcal{S}}_{\alpha,\beta}
     \nonumber\\
     &=&
    \sum_{m=-\infty}^\infty
    c_m^{\rm BD}\left(\cosh\alpha\,a^{\dagger}_m+
       e^{-i\beta}(-1)^m \frac{\gamma_m^*}{\gamma_m}\sinh\alpha\,a_m\right)\,+\,{\rm h.c.}
 \ea
From eq.(\ref{cBD}), we have the relation between general coefficients $c_m$ and $c_m^{\rm{BD}}$:
\be
       c_m=c_m^{\mathrm{BD}}(\cosh\alpha+(-1)^m e^{i\beta}\sinh\alpha)\,.
\ee
 This implies that
 \be
 	 {\cal S}_{\alpha,\beta}^{\dagger}\,\Phi_{\rm BD}(0)\,{\cal S}_{\alpha,\beta}=
	 \Phi_{\alpha,\beta}(0)\,.
\ee
Moreover since the squeezing operator  commutes with all generators, we have, 
at every space-time point, 
\be
       {\mathcal{S}}_{\alpha,\beta}^{\dagger}\,\Phi_{\rm{BD}}(x)\,{\mathcal{S}}_{\alpha,\beta}
       \,=\,\Phi_{\alpha,\beta}(x)\,.
\ee
This establishes in full generality
the relation between the usual QFT treatment and our approach.
Indeed, collecting the various results, we have
\ba
    G_{\alpha,\beta}(x,x')&=&\langle
   \Omega|\,\Phi_{\alpha,\beta}(x)\,\Phi_{\alpha,\beta}(x')\,|\Omega\rangle,\nonumber\\
   &=&\langle
   \Omega|\,{\mathcal{S}}_{\alpha,\beta}^\dagger\,\Phi_{\rm{BD}}(x)\,\Phi_{\rm{BD}}(x')\,{\mathcal{S}}_{\alpha,\beta}
   |\Omega\rangle,\nonumber\\
   &=&\langle \Omega_{\alpha,\beta}|\,\Phi_{\rm{BD}}(x)\,\Phi_{\rm{BD}}(x')\,
   |\Omega_{\alpha,\beta}\rangle\,, 
\ea
where $|\Omega_{\alpha,\beta}\rangle$ is the $\alpha$-vacuum
conventionally defined.

\section{Arbitrary dimensions}

In this section we shall show that 
the approach we have used in two dimensions
can be generalized to arbitrary dimensions. In particular,
when considering the principal series, that is, for scalar fields with mass squared 
greater than $(n-1)^2/4$, 
we shall see that the moduli space of covariant canonical fields
stays SU$(1,1)/$U$(1)$ irrespectively of the dimensionality.  On the contrary the
{\it in}-{\it out} Bogoliubov coefficients do depend on the dimensionality:
they vanish for all odd dimensional dS spaces. 

The $n$-dimensional de Sitter space, $dS_n$ is described by the hyperboloid in $(n+1)$-dimensional
Minkowski space ${\mathbb{R}}^{1,n}$:
\be
   \eta_{AB} X^A X^B\,=\,1\,,
   \label{dSn}
\ee
where $\eta_{AB}=diag(-1,1,\dots ,1)$. In the following we use the index notation:
\be
\begin{array}{ll}
       A,B,C,D=0,1,\dots ,n\,; &\quad I,J=1,2,\dots ,n\,;\\
       \mu,\nu=0,1,\dots ,n-1\,;&\quad i,j,k=1,2,\dots ,n-1\,.\\
\end{array}
\ee

\subsection{The SO$_0(1,n)$ group}

The isometry group of $dS_n$ is SO$_0(1,n)$. It
 is the group of transformations continuously connected to the identity 
 which leaves eq.(\ref{dSn}) invariant. The generators of SO$_0(1,n)$  verify the following algebra:
\be
       [\,{\cal M}_{AB},\,{\cal M}_{CD}\,]\,=\,-i\,\left(\,
       \eta_{AC}{\cal M}_{BD}-\eta_{AD}{\cal M}_{BC}-\eta_{BC}{\cal M}_{AD}+\eta_{BD}{\cal M}_{AC}
       \,\right).
\ee
When considering the principal series, the 
UIR of  SO$_0(1,n)$ can be realized by the square integrable functions
on $S^{n-1}$ : ${\cal L}^2(S^{n-1})$. The $(n-1)$-sphere is 
conventionally parameterized by a vector $\vec\zeta$ in $\mathbb{R}^n$
 subject to $|\vec\zeta|=1$. 
 The action of the generators of SO$_0(1,n)$ on the UIR states are given by
\ba
   {\cal M}_{IJ}
   &=&i\left(\zeta_I{\partial \over \partial \zeta^J}-
   \zeta_J{\partial \over \partial \zeta^I}\right)\,,
   \nonumber\\
   {\cal M}_{I0}
   &=&{1\over 2}\left({\cal M}_{IJ}\zeta^J+\zeta^J{\cal M}_{IJ}\right)+\mu\,\zeta_I
   \nonumber\\
   &=&\zeta^J{\cal M}_{IJ}
   +\left(\mu+i{n-1\over 2}\right)\zeta_I\,,
   \label{dSgen}
\ea
and the quadratic Casimir is given by
\be
    {\cal C}\,=\,\frac12\sum_{A,B}{\cal M}_{AB}{\cal M}^{AB}\,=\,-\mu^2-\frac{(n-1)^2}2\,.
\ee
It would be useful to have the action under finite transformations, we have
\ba
   &\langle\,\vec\zeta\,\vert \,e^{i\theta{\cal M}^{IJ}}\,\vert \Psi\rangle
   &=\,\langle\,\vec\zeta'\,\vert \Psi\rangle\,,
   \nonumber\\
   &\langle\,\vec\zeta\,\vert \,e^{i\omega{\cal M}^{I0}}\,\vert \Psi\rangle
   &=\,(\cosh\omega+\zeta^I\sinh\omega)^{i\mu-{n-1\over 2}}
   \langle\,\vec\zeta''\,\vert \Psi\rangle\,,
   \label{fdSact}
\ea
where
\ba
	&&\vec\zeta'=e^{i\theta{\mathscr M}^{IJ}}\vec\zeta\,,
	\nonumber\\
	&&\vec\zeta''=\left({\zeta^1\over \cosh\omega+\zeta^I\sinh\omega},\,\dots\,,
   	{\sinh\omega+\zeta^I\cosh\omega\over\cosh\omega+\zeta^I\sinh\omega},\,
   	\dots\,,{\zeta^n\over \cosh\omega+\zeta^I\sinh\omega}\right),
	\label{fdSactw}
\ea
and where ${\mathscr M}^{IJ}$ is the representation of ${\cal M}^{IJ}$ on the vector space ${\mathbb R}^n$.
The inner product of two position eigenstates is
\be
   \langle\,\vec\zeta\,\vert\,\vec\zeta'\,\rangle=\delta^{n-1}(\vec\zeta-\vec\zeta')\,,
\ee
where the delta function $\delta^{n-1}(\vec\zeta)$ is defined on $S^{n-1}$.

\subsection{Massive scalar field in $dS_n$}

We choose the origin in $dS_n$ to have the embedding coordinates $X_0^A=(0,0,\dots ,0,1)$.
The field on any point of $dS_n$, of coordinates  $X^A=\Lambda^A_{\phantom{A}B} X_0^B$,
 can be deduced from the field at the origin $\Phi(X_0)$ by
\be
   \Phi(X)=U(\Lambda) \, \Phi(X_0) \, U(\Lambda)^{-1}\,,
   \label{fondeq}
\ee
where $\Lambda$ is an element of SO$_0(1,n)$.

In the global coordinate system:
\ba
   X^0&=&\sinh t\,,\nonumber\\
   X^I&=&\cosh t\,\xi^I\,,\qquad\vec\xi\in S^{n-1}\,,\nonumber\\
   \xi^I &=&{\mathrm{R}}^I_{\phantom{I}J}\,\xi_0^J\,,\qquad\vec\xi_0=(0,\dots,0,1)\,,
\ea
where $\mathrm{R}$ is a element of SO$(n)$ subgroup. The metric reads
\be
   ds^2=-dt^2+\cosh^2 t\, d\Omega^2(\vec\xi)\,.
\ee
Therefore the point $X_0$ can be transported to any point $X$ by a boost followed by a rotation.
This implies that eq.(\ref{fondeq}) can be written as
\be
   \Phi(X)=U({\mathrm{R}})\, e^{itM^{0n}}\,\Phi(X_0)\,e^{-itM^{0n}}\,U({\mathrm{R}})^{-1}\,.
\ee

The origin $X_0=(0,\vec\xi_0)$ is invariant under the action of the subgroup
SO$_0(1,n-1)$ generated by $M_{\mu\nu}$. To construct a local field, we thus require
\be
   [\,M_{\mu\nu},\,\Phi(0,\vec\xi_0)\,]=0\,.
   \label{cvn}
\ee
As in two dimensions, after some algebra, one sees that 
the Casimir relation gives the Klein-Gordon equation:
\be
    \left(\partial_t^2+(n-1)\tanh t\,\partial_t-\frac1{\cosh^2t}\,\Delta_{\vec\xi}\right)
    \Phi(t,\vec\xi)=-\left(\mu^2+\frac{(n-1)^2}4\right)\Phi(t,\vec\xi)\,,
\ee
where $\Delta_{\vec\xi}$ is the Laplacian on $S^{n-1}$.

From the UIR of SO$_0(1,n)$ we define the creation and annihilation operators by
\ba
   &a^\dagger(\vec\zeta)\vert\Omega\rangle=\vert\,\vec\zeta\,\rangle\,,
   \nonumber\\
   &[\,a(\vec\zeta),\,a^\dagger(\vec\zeta')\,]=\delta^{n-1}(\vec\zeta-\vec\zeta')\,,
   \qquad [\,a(\vec\zeta),\,a(\vec\zeta')\,]=0\,.
\ea
The field operator in the origin can be expanded in terms of these operators:
\be
   \Phi(0,\vec\xi_0)=\int d^{n-1}\Omega(\vec\zeta)\
  \left[ \Psi_0(\vec \zeta)\,a^\dagger(\vec\zeta)\,+\,\Psi_0^*(\vec\zeta)\,a(\vec\zeta)
  \right],
\ee
where $d^{n-1}\Omega(\vec\zeta)$ is the invariant volume element on $S^{n-1}$.

The covariance condition (\ref{cvn}) determines
 the function $\Psi_0(\vec\zeta)$. The rotation part of this equation implies
 that $\Psi_0(\vec\zeta)$ depends on $\vec\zeta$ only through $\vec\zeta\cdot\vec\xi_0$,
 whereas the boost part fixes this dependence to be 
 again governed by two arbitrary coefficients:
\ba
   \Psi_0(\vec\zeta)&=&A\,\Theta(\vec\zeta\cdot\vec\xi_0)\,\left(\vec\zeta\cdot\vec\xi_0\right)^{i\mu-{n-1\over 2}}
   \,+\,B\,\Theta(-\vec\zeta\cdot\vec\xi_0)\,\left(-\vec\zeta\cdot\vec\xi_0\right)^{i\mu-{n-1\over 2}}
   \nonumber\\
   &=&C\,(\vec\zeta\cdot\vec\xi_0-i\epsilon)^{i\mu-\frac{n-1}2}
   +D\,(\vec\zeta\cdot\vec\xi_0+i\epsilon)^{i\mu-\frac{n-1}2}\,,
   \label{ffnn}
\ea
where the limit $\epsilon\to 0^+$ is understood. We have also the following relation between the coefficients:
\be
   A=C+D\,,\qquad B=(e^{-i\pi})^{i\mu-{n-1\over 2}}\,C+(e^{i\pi})^{i\mu-{n-1\over 2}}\, D\,,
\ee
which generalize eq.(\ref{ABCD}).
Transporting the field with a boost followed by a rotation
brings the point $(0,\vec\xi_0)$ to
$(t,\vec\xi)$. We thus get
\be
   \Phi(t,\vec\xi)\,=\,\int d^{n-1}\Omega(\vec\zeta)\ \Psi_{t,\vec\xi}(\vec\zeta)\,a^\dagger(\vec\zeta)
   \,+\, \Psi_{t,\vec\xi}^*(\vec\zeta)\,a(\vec\zeta)\,,
\ee
with
\be
   \Psi_{t,\vec\xi}(\vec\zeta)\,=\,C\,(\cosh t\,\vec\zeta\cdot\vec\xi+\sinh t-i\epsilon)^{i\mu-\frac{n-1}2}
       +D\,(\cosh t\,\vec\zeta\cdot\vec\xi+\sinh t+i\epsilon)^{i\mu-\frac{n-1}2}\,.
      \label{CD}
\ee
The two point function is simply expressed in terms of
$\Psi_{t,\xi}(\vec\zeta)$ as
\be
   \langle\Omega\vert\,\Phi(t,\vec\xi)\,\Phi(t',\vec\xi')\,\vert\Omega\rangle\,=\,\int d^{n-1}\Omega(\vec\zeta)\
   \Psi_{t,\vec\xi}^*(\vec\zeta)\,\Psi_{t',\vec\xi'}(\vec\zeta)\,.
   \label{ntpf}
\ee
The above integral can be expressed in terms of hypergeometric functions as shown in the Appendix A.
We obtain
\ba
   \langle\Omega|\,\Phi(x)\,\Phi(x')\,|\Omega\rangle
   &=&|C|^2e^{\pi\mu}\,F_n(x;x')\,+\,|D|^2e^{-\pi\mu}\,
   F_n(\bar x;\bar x')
   \nonumber\\
   &&+\,2\,{\rm Re}\left[C^*D\,e^{-i\frac{n-1}2\pi}\,F_n(x;\bar x')\right],
   \label{dSWn}
\ea
with
\be
   F_n(x;x')=\frac{2\pi^{\frac n2}}{\Gamma(\frac n2)}\,
   {}_2F_1\left(i\mu+\frac{n-1}2,-i\mu+\frac{n-1}2;\frac n2;\frac{1+\tilde Z(x;x')}2\right).
\ee
We have defined $\tilde Z$ by
\be
	\tilde Z(x;x')=Z(x;x')+i\,{\rm sgn}(t-t')\,\epsilon\,,
	\label{tildeZ}
\ee
where $Z$ is the dS invariant quantity:
\be
	Z(t,\vec\xi;t',\vec\xi')=
	\cosh t\cosh t'\,\vec\xi\cdot\vec\xi'-
   	\sinh t\sinh t'=X^AX'_A\,.
\ee

Because of the $i \epsilon$ term, $\tilde Z$ is not a dS invariant for $\vert Z\vert<1$. 
Nevertheless the two point  function is dS invariant. The reason is the following. 
  The hypergeometric function has a branch cut for $Z>1$.  
Therefore the $i\epsilon$ term is irrelevant for $Z<1$. When $Z>1$ instead, 
it becomes relevant but ${\rm sgn}(t-t')={\rm sgn}(X^0-X'^0)$ 
and hence $\tilde Z$ are dS invariant.
It should be also emphasized that the $i\epsilon$ term in eq.(\ref{dSWn})
originates from eq.(\ref{ffnn}).
Therefore what we obtain is the positive Wightman
function.

As in the two dimensional case, the reality of the two point function for space-like separation,
or the parity argument, insures that the equal time commutator of two fields vanishes.

Proceeding as in Section 4, it can be easily seen that the canonical
commutation relation is satisfied modulo a constant which should be
identified with $\hbar$. More explicitly we have
\be
   \left[\,\partial_t \Phi(0,\vec\xi_0),\,\Phi(0,\vec\xi)\,\right]=iN_n\,\delta^{n-1}(\vec\xi_0-\vec\xi)\,,
\ee
with
\be
   N_n= 2\,{\rm Im}\left[\int d^{n-1}\Omega(\vec \xi) \int d^{n-1}\Omega(\vec \zeta)\;
   \partial_t\Psi_{0,\vec\xi_0}^*(\vec\zeta)\,
   \Psi_{0,\vec\xi}(\vec\zeta)\right].
\ee
Using the explicit form of $\Psi_{t,\vec\xi}(\vec\zeta)$, the integral can be calculated, and the
canonical commutation relation leads to
\be
   -e^{\pi\mu}|C|^2+e^{-\pi\mu}|D|^2=2^{-n-1}\pi^{-n}\left|\,
   \Gamma\left({i\mu+\frac{n-1}2}\right)\,\right|^2\,.
   \label{ccrn}
\ee

By defining the vector:
\be
   {\bf{z}}=\frac{2^{\frac{n+1}2}\pi^{\frac{n}2}}
   {\left|\Gamma\left({i\mu+\frac{n-1}2}\right)\right|}\,
   \binom{e^{-\frac{\pi\mu}2}D}{e^{\frac{\pi\mu}2}e^{i\frac{n-1}2\pi}C}\,,
\ee
the canonical condition implies eq.(\ref{ccrz}), 
therefore its solutions are still of the form eq.(\ref{paraz}).

If we now impose the time reversal symmetry on the field, ${\mathsf{T}}\Phi(t,\vec\xi){\mathsf{T}}^{-1}=\Phi(-t,\vec\xi)$,
we get a constraint for the coefficients $C$ and $D$.

As in the two dimensional case, the {\it unitary} time reversal operator is not compatible with the canonical
commutation relation ({\ref{ccrn}}), since it gives
\be
   e^{\pi\mu}\,\vert C\vert ^2\,=\,e^{-\pi\mu}\,\vert D\vert ^2\,.
\ee

With the {\it{anti-unitary}} time reversal operator instead, we get
\ba
   \langle\Omega\vert\,\Phi(t,\vec\xi)\,\Phi(t',\vec\xi')\,\vert\Omega\rangle
   &=&\langle\Omega\vert{\mathsf{T}}^{-1}\,
   {\mathsf{T}}\Phi(t,\vec\xi){\mathsf{T}}^{-1}\,
   {\mathsf{T}}\Phi(t',\vec\xi'){\mathsf{T}}^{-1}\,
   {\mathsf{T}}\vert\Omega\rangle
   \nonumber\\
   &=&\langle\Omega\vert\,\Phi(-t,\vec\xi)\,\Phi(-t',\vec\xi')\,\vert\Omega\rangle^*\,.
\ea
This gives a constraint which is compatible with the canonical commutation relation:
\be
   e^{-i\frac{n-1}2\pi}\,C^*\,D\,=\,e^{i\frac{n-1}2\pi}\,C\,D^*\,.
   \label{trn}
\ee
In terms of the vector $\bf z$, the above relation reads 
\be
   {\bf z}^\dagger\sigma_2 \,{\bf z}=0\,,
\ee
which implies $\beta=0$, as before.

\subsection{Bunch-Davies vacuum and {\it in} and {\it out} vacua}

In the coincidence point limit,
the Minkowski vacuum positive Wightman function behaves as
\be
   G_{\rm{Mink}}(x;x')\underset{ x\to x'}{\approx}
   \frac1{4\pi^{n\over 2}}\times
   \left\{
   \begin{array}{ll}
       \log|\,(x-x')^2-(t-t'-i\epsilon)^2\,|^{-2}\,, &\;{\rm for}\; n=2\,,\\
       &\\
       \Gamma({n\over 2}-1)
       \vert\,(\vec x-\vec x')^2-(t-t'-i\epsilon)^2\,\vert^\frac{2-n}2\,, &\;{\rm for}\; n\neq2\,.
   \end{array} \right.
\ee
The short distance behavior of two point function in $dS_n$, eq.(\ref{dSWn}), is
\be
   G_{dS_n}(x;x')\underset{ x\to x'}{\approx}
   \frac{\pi^{\frac n2}2^{n-1}}{\left|\Gamma\left(i\mu+\frac{n-1}2\right)\right|^2}\times
   \left\{
   \begin{array}{ll}
       |C|^2e^{\pi\mu}\,\log|\,(x-x')^2-(t-t'+i\epsilon)^2\,|^{-2} &\\
       +\,|D|^2e^{-\pi\mu}\,\log|\,(x-x')^2-(t-t'-i\epsilon)^2\,|^{-2}\,,&\;{\rm for}\;n=2\,,\\
       &\\
       |C|^2e^{\pi\mu}\,\vert\,(\vec x-\vec x')^2-(t-t'+i\epsilon)^2\,\vert^\frac{2-n}2&\\
       +\,|D|^2e^{-\pi\mu}\,\vert\,(\vec x-\vec x')^2-(t-t'-i\epsilon)^2\,\vert^\frac{2-n}2\,,&\;{\rm for}\;n\neq2\,.
   \end{array} \right.
\ee
 This has been obtained from the asymptotic behavior of the hypergeometric function 
 presented in Appendix A.
 Imposing that the behavior be that of Hadamard, one gets the Bunch-Davies coefficients:
\be
   C^{\mathrm{BD}}=0\,,\qquad D^{\mathrm{BD}}=\frac{e^\frac{\pi\mu}2}{2^\frac{n+1}2\pi^\frac{n}2}
   \left|\,\Gamma\left({i\mu+\frac{n-1}2}\right)\,\right|.
\ee
In terms of ${\bf{z}}$, the Bunch-Davies vacuum still corresponds to ${\bf{z^{{\rm{BD}}}}}=\binom10$, and
the general ($C,D$) vacuum can be obtained from the Bunch-Davies coefficients:
\be
       C=e^{i\beta}\sinh\alpha\, e^{-\pi(\mu+i\frac{n-1}2)}\,D^{\rm{BD}}\,,\qquad
       D=\cosh\alpha\, D^{\rm{BD}}\,.
\ee

The behavior of the field in asymptotic past and future 
can be extracted from that of the two point function.
We put a field operator in the far past (or future) with coefficients $C$ and $D$ and another field at the origin
with different coefficients $\tilde C$ and $\tilde D$ to avoid any 
confusion. We thus analyze
\be
   \langle\Omega|\,\Phi_{\tilde C,\tilde D}(0,\vec{\xi_0})\,\Phi_{C,D}(t,\vec\xi_0)\,|\Omega\rangle\,,
\ee
in the limit $t\to\pm\infty$.
From the asymptotic expression of the hypergeometric function 
given in Appendix A, we can see that the two
point function has the following behavior:
\ba
   &&\langle\Omega|\,\Phi_{\tilde C,\tilde D}(0,\vec{\xi_0})\,\Phi_{C,D}(t,\vec\xi_0)\,|\Omega\rangle
   \underset{t\to\pm\infty}{\approx}
   \frac{2^n\pi^{\frac n2}}
   {\left|\Gamma\left(i\mu+\frac{n-1}2\right)\right|^2}\times
   \nonumber\\
   &&\times\Bigg[\,
   2^{2i\mu}\Gamma(-2i\mu)
   \Big\{
   \tilde{C}^*C\,e^{\pi\mu}\,(-e^{\pm t}\pm i\epsilon)^{-i\mu-\frac{n-1}2}
   +\,\tilde{D}^*D\,e^{-\pi\mu}\,(-e^{\pm t}\mp i\epsilon)^{-i\mu-\frac{n-1}2}
   \nonumber\\
   &&\qquad\qquad\qquad\quad
   +\,\tilde C^*D\,e^{-i\frac{n-1}2\pi}\,(e^{\pm t}\mp i\epsilon)^{-i\mu-\frac{n-1}2}
   +\,\tilde D^*C\,e^{i\frac{n-1}2\pi}\,(e^{\pm t}\pm i\epsilon)^{-i\mu-\frac{n-1}2}
   \Big\}
   \nonumber\\
   &&\qquad+\,2^{-2i\mu}\Gamma(2i\mu)
   \Big\{
   \tilde C^*C\,e^{\pi\mu}\,(-e^{\pm t}\pm i\epsilon)^{i\mu-\frac{n-1}2}
   +\,\tilde D^*D\,e^{-\pi\mu}\,(-e^{\pm t}\mp i\epsilon)^{i\mu-\frac{n-1}2}
   \nonumber\\
   &&\qquad\qquad\qquad\qquad
   +\,\tilde C^*D\,e^{-i\frac{n-1}2\pi}\,(e^{\pm t}\mp i\epsilon)^{i\mu-\frac{n-1}2}
   +\,\tilde D^*C\,e^{i\frac{n-1}2\pi}\,(e^{\pm t}\pm i\epsilon)^{i\mu-\frac{n-1}2}
   \Big\}\,\Bigg]\,.\nonumber\\
\ea
In remote past we have a negative frequency term proportional to
\be
   \left(\,e^{\pi\mu}\,\tilde C^*+e^{-\pi\mu}\,\tilde D^*\,\right)
   \left(\,e^{\pi(\mu+i\frac{n-1}2)}\,C+\,e^{-\pi(\mu+i\frac{n-1}2)}\,D\,\right)\,e^{-i\mu t}\,.
\ee
The {\it in} vacuum thus corresponds to 
\be
   e^{\pi(\mu+i\frac{n-1}2)}\,C^{\rm{IN}}+\,e^{-\pi(\mu+i\frac{n-1}2)}\,D^{\rm{IN}}=0\,.
\ee
Similarly, in the far future we have
\be
   \left(\,e^{\pi(\mu-i\frac{n-1}2)}\,\tilde C^*+\,e^{-\pi(\mu-i\frac{n-1}2)}\,\tilde D^*\,\right)
   \Big(\,e^{\pi\mu}\,C+e^{-\pi\mu}\,D\,\Big)\,e^{-i\mu t}\,,
\ee
and the {\it out} vacuum corresponds to
\be
   e^{\pi\mu}\,C^{\rm{OUT}}+\,e^{-\pi\mu}\,D^{\rm{OUT}}=0\,.
\ee
Notice that for odd dimensions, we have the same condition for the {\it in} and 
the {\it out} vacuum.
This implies that there is no particle creation in odd dimensions
for all scalar fields belonging to the principal series.
It will be interesting to further investigate this
phenomenon. 

In terms of $A$ and $B$ the above conditions read 
\ba
   &B^{\rm{IN}}\,=\,0\,,
   \nonumber\\
   &\sin(\frac{n-1}2\pi)\,A^{\rm{OUT}}\,=\,i\,\sinh\pi\mu\,B^{\rm{OUT}}\,,
\ea
 which clearly show that $B^{\rm{IN}}=B^{\rm{OUT}}=0$
 when $n$ is odd.
 As in two dimensions therefore we find that the {\it in} vacuum is characterized by 
a state $\vert\Psi_0\rangle$ in the UIR 
whose support is inside the past horizon of $(t=-\infty,\vec\xi_0)$, see eq.(\ref{ffnn}).

\acknowledgments{
We are grateful to X. Bekaert, E. Buffenoir, J.P. Gazeau, K. Noui, and
 Ph. Roche for stimulating discussions.}
 
\appendix

\section{Wightman function in arbitrary dimension}

In this Appendix we compute the Wightman function in $n$-dimensions.
In the treatment we adopted, the $i\epsilon$ prescription follows from the 
writing of the vector state $\vert \Psi_0\rangle$
in terms of the holomorphic and anti-holomorphic sectors weighted by the coefficients $C$ and $D$
respectively, see eq.(\ref{CD}). 
Instead, in the usual approach of quantum fields in dS, see \cite{Allen:1985ux}, 
the $i\epsilon$ prescription  is ambiguous: one obtains 
the Wightman function by first solving the Klein-Gordon equation assuming that it
depends only on the invariant quantity $Z=\eta_{AB}X^A Y^B$. 
The ambiguity arises because $Z$ is invariant under the exchange
of the two points and because  Wightman functions are sensitive to the
ordering for time-like separated points.
 In order to cure this problem, one introduces from the outset an imaginary 
prescription (an  $i\epsilon$ term) to $Z$, the sign of which is that of 
$X^0-X'^0$, see eq.(\ref{tildeZ}).

As explained in section 8, the two point function in $n$-dimensions can be obtained as
\be
   G(t,\vec{\xi};t',\vec{\xi'})\,=\,\langle\Omega|\,\Phi(t,\vec{\xi})\,\Phi(t',\vec{\xi'})\,|\Omega\rangle
   \,=\,\int d^{n-1}\Omega(\vec{\zeta})\,\Psi^*_{t,\vec\xi}(\vec\zeta)\Psi_{t',\vec\xi'}(\vec\zeta)\,,
\ee
with $\Psi_{t,\vec\xi}(\vec\zeta)$ given in eq.(\ref{CD}).
Each term with coefficients $|C|^2$, $|D|^2$, $C^*D$ and $D^*C$ can be written in terms of a 
 single function $f$ with appropriate arguments as
\ba
   G(x;x')&=&|C|^2\,f(x;x')\,+\,|D|^2e^{-2\pi\mu}\,
   f(\bar x;\bar x')\nonumber\\
   &&+\,C^*D\,e^{-\pi\mu}e^{-i\frac{n-1}2\pi}\,f(x;\bar x')
   \,+\,D^*Ce^{-\pi\mu}e^{i\frac{n-1}2\pi}\,f(\bar x;x')\,,
   \label{Gf}
\ea
with 
\be
   f(t,\vec\xi;t',\vec\xi')=\int d^{n-1}\Omega(\vec{\zeta})\,
   (\cosh t\,\vec\zeta\cdot\vec\xi+\sinh t+i\epsilon)^{-i\mu-\frac{n-1}2}\,
   (\cosh t'\,\vec\zeta\cdot\vec\xi'+\sinh t'-i\epsilon)^{ i\mu-\frac{n-1}2}\,.
   \label{intf}
\ee
Using the following identity, valid for positive $\epsilon$:
\be
   \frac1{(A+i\epsilon)^\alpha
   (B-i\epsilon)^\beta}=e^{i\pi\beta}\frac{\Gamma(\alpha+\beta)}{\Gamma(\alpha)\Gamma(\beta)}
   \int_0^1dx\int_0^1dy\,\delta(x+y-1)
   \frac{x^{\alpha-1}y^{\beta-1}}{(xA-yB+i\epsilon)^{\alpha+\beta}}\,,
\ee
the integral in eq.(\ref{intf}) can be put in the form:
\ba
   &&\frac{e^{\pi\mu}\,e^{i\frac{n-1}2\pi}\,\Gamma(n-1)}{\Gamma\left(i\mu+\frac{n-1}2\right)
   \Gamma\left(-i\mu+\frac{n-1}2\right)}
   \int_0^1dx\int_0^1dy\,\delta(x+y-1)\,x^{-i\mu+\frac{n-3}2} y^{i\mu+\frac{n-3}2} \nonumber\\
   &&\;\times\int d^{n-1}\Omega(\vec{\zeta})\,\Big(\,\vec\zeta\cdot(x\cosh t\,
   \vec\xi-y\cosh t'\,\vec\xi')
   +(x\sinh t-y\sinh t')+i\epsilon\Big)^{1-n}\,.
\ea
Introducing $\vec a$ and $b$ by
\be
   \vec a=x\cosh t\,\vec\xi-y\cosh t'\,\vec\xi'\,,\qquad
   b=x\sinh t-y\sinh t'\,,
\ee
and defining $\phi$ as the angle between the vector $\vec\zeta$ and $\vec a$,
the integral with respect to $\vec\zeta$ on $S^{n-1}$ reads
\be
   \int_0^\pi d\phi\,\sin^{n-2}\phi\,
   \Big( \vert \vec a \vert \cos\phi+b+i\epsilon\Big)^{1-n}\,\int d^{n-2}\Omega\,,
\ee
where the integral over $S^{n-2}$ gives the area
${2\pi^{\frac{n-1}2}}/{\Gamma(\frac{n-1}2)}$. The remaining integral can be calculated as
\ba
   &\int_0^\pi d\phi\,\sin^{n-2}\phi\,(\vert \vec a \vert\cos\phi+b+i\epsilon)^{1-n}
   \,=\,(-1)^{n-1}\int_0^\pi
   d\phi\,\sin^{n-2}\phi\,\Big( \vert \vec a \vert +(b+i\epsilon)\cos\phi\,\Big)^{1-n}
   \nonumber\\
   &=\,(-1)^{n-1}\pi^{1/2}\frac{\Gamma(\frac{n-1}2)}
   {\Gamma(\frac n2)}(\vert \vec a \vert^2-b^2-2\,i\epsilon\,b)^{\frac{1-n}2}
   =\,(-1)^{n-1}\pi^{1/2}\frac{\Gamma(\frac{n-1}2)}{\Gamma(\frac n2)}
   (x^2+y^2-2xy\,\tilde Z)^{\frac{1-n}2}\,,\nonumber\\
\ea
where $\tilde Z$ is the time ordered invariant distance:
\be
   \tilde Z(t,\vec\xi;t',\vec\xi')=\cosh t\cosh t'\,\vec\xi\cdot\vec\xi'-
   \sinh t\sinh t'+i\,\mathrm{sgn}(t-t')\,\epsilon\,.
\ee
Finally, the integral with respect to $x$ and $y$ gives a hypergeometric function:
\ba
   f&=&e^{\pi\mu}\frac{2\pi^{\frac n2}}{\Gamma(\frac n2)}
   \frac{\Gamma(n-1)}{\Gamma\left(i\mu+\frac{n-1}2\right)
   \Gamma\left(-i\mu+\frac{n-1}2\right)}
   \nonumber \\ 
   && \quad \times \int_0^1dx\int_0^1dy\,\delta(x+y-1)\,x^{-i\mu+\frac{n-3}2}
   y^{i\mu+\frac{n-3}2}(x^2+y^2-2xy\tilde Z)^{1-n} , 
   \nonumber\\
   &=&e^{\pi\mu}\frac{2\pi^{\frac n2}}{\Gamma(\frac n2)}
   \,{}_2F_1\left(i\mu+\frac{n-1}2,-i\mu+\frac{n-1}2;\frac n2;\frac{1+\tilde Z}2\right).
\ea
If we now substitute in eq.(\ref{Gf}), the obtained expression of $f$ we get eq.(\ref{dSWn}) used in the text.

To conclude this Appendix, we present two formula used in section 8.
The small distance behavior of the two-point is governed by 
the limit given in \cite{bateman}:
\be
   {}_2F_1\left(i\mu+{{n-1}\over 2},
   -i\mu+{n-1\over 2}; {n\over 2};\frac{1+x}2\right)
   \underset{x\to1}{\approx}\left\{
   \begin{array}{ll}
       \frac{-2}{\left|\Gamma\left(i\mu+\frac12\right)\right|^2}\,\log(1-x)\,, &\;{\rm for}\;n=2\,,\\
       &\\
       \frac{2^{n-2}\,\Gamma\left(\frac n2\right)\Gamma\left(\frac n2-1\right)}
       {\left|\Gamma\left(i\mu+\frac{n-1}2\right)\right|^2}\,(1-x)^{2-n}\,, &\;{\rm for}\; n\neq2\,.
   \end{array} \right.
\ee
In subsection 8.3, we used the asymptotic behavior 
of the two-point. It is governed by \cite{bateman}:
\ba
   &&{}_2F_1\left(i\mu+\frac{n-1}2,-i\mu+\frac{n-1}2;\frac n2;\frac{1+x}2\right)
   \nonumber\\
   &&\qquad\underset{x\to\infty}{\approx}
   \frac{2^\frac{n-2}2\Gamma\left(\frac{n}2\right)}
   {\left|\Gamma\left(i\mu+\frac{n-1}2\right)\right|^2}
   \left[\,2^{i\mu}\Gamma(-2i\mu)\,(-x)^{-i\mu-\frac{n-1}2}
   +2^{-i\mu}\Gamma(2i\mu)\,(-x)^{i\mu-\frac{n-1}2}\,\right].\nonumber\\
\ea

\section{Flat sections}

In this Appendix we consider the quantization
of a massive scalar field using flat sections.
This amounts to use the basis of the UIR which is
adapted to the isometry group of these.
Since the state $\vert{\Psi_0}\rangle$ is unchanged, 
its components in the new basis
give the expansion of the field operator in terms of the new 
creation and annihilation operators. 

 We first recall the  flat section parameterization:
 \ba
   X^0&=&-\sinh\tau-\frac12e^{\tau}\sum_{i=1}^{n-1} ( x^i)^2\,,\nonumber\\
   X^i&=&e^\tau x^i\,, \nonumber\\
   X^n&=&\cosh\tau-\frac12 e^{\tau}\sum_{i=1}^{n-1} ( x^i)^2\,.  
 \ea
 It covers the $X^n>X^0$ part of dS space and 
 leads to the metric:
 \be
 	ds^2=-d\tau^2+e^{2\tau}\sum_{i=1}^{n-1} (d x^i)^2\,,
\ee
showing that the isometry group of the constant $\tau$ sections is
 the $(n-1)$-dimensional Euclidean group.
 The infinitesimal action of the Killing vector fields is given by
 \be
 \begin{array}{llll}
   {\cal M}_{n0}&:&\delta \tau=\epsilon\,,&\delta \vec x= - \epsilon \vec x \,,\\
   {\cal P}_i=\frac{1}{\sqrt{2}}({\cal M}_{i0}-{\cal M}_{in}) &:&\delta\tau=0\,,&\delta x^i=\epsilon \,, \\
   {\cal M}_{ij}&:&\delta \tau=0\,,&\delta x^i=\epsilon\,, \ \delta x^j=-\epsilon\,,\\
   \frac{1}{\sqrt{2}}({\cal M}_{i0}+{\cal M}_{in})&:&\delta\tau=\epsilon x^i\,,
   &\delta x^i=-\frac{\epsilon}{2}\left( (x^i)^2+e^{-2\tau}\right).
 \end{array}
 \ee
 These allow to identify ${\cal P}_i$ as the translation generators.
 The point with coordinates $(0,\vec 0)$ is left invariant by
 the infinitesimal transformation generated by $M_{\mu\nu}$.
 The transformation generated by the boost $M_{n0}$ followed
 by a one generated by the translations $P_i$ allows to reach an
 arbitrary point
 with coordinates $(\tau,\vec x)$. So we now  have
 \be
   \Phi(\tau,\vec x)=  e^{-i\vec x\cdot \vec P} e^{i\tau M_{n0}}
   \Phi(0,\vec 0) e^{-i\tau M_{n0}}e^{i\vec x\cdot\vec P}\,,
   \label{sep}
 \ee
 with
 \be
   [\,M_{\mu\nu},\,\Phi(0,\vec0)\,]=0\,.
 \ee
 We have already solved this equation in eq.(\ref{ffnn}).
 A convenient basis of the UIR  is formed
 by the eigenvectors $|\,\vec p\,\rangle $ of $P_i$. 
 The action of the translations is diagonal on this basis.
 The explicit writing of eq.(\ref{sep}) amounts to calculate $\langle\, \vec p\,|\Psi_0\rangle $
 and $e^{i\tau M_{n0}}|\,\vec p\,\rangle $.
 Since we have already the decomposition of $|\Psi_0\rangle $ on the position eigenstates $|\,\vec \zeta\,\rangle $
it is sufficient to calculate the connector $\langle \,\vec \zeta\,|\,\vec p\,\rangle $.  Using the expression of the generators given in eq.(\ref{dSgen}), a direct calculation gives
\be	
	\langle\,\vec\zeta\,\vert\,\vec p\,\rangle=
	2^{-\frac n2} \pi^{-\frac{n-1}2}\,(1-\zeta_n)^{i\mu-\frac{n-1}2}
	\exp\left(i\frac{\vec p\cdot\vec\zeta'}{1-\zeta_n}\right).
	\label{pp}
\ee
In this equation, we have introduced a new notation.
The vector ${\vec \zeta}$ lives on the $(n-1)$-dimensional sphere. 
It can thus be described by the following $n$ coordinates $(\vec \zeta', \zeta_n)$
where $\vec \zeta'$ lives, like $\vec p$, in a $(n-1)$-dimensional Euclidean space.

The action of the boost $e^{i\tau M_{n0}}$ on $|\,\vec p\,\rangle $ can also be calculated using 
eq.(\ref{fdSact}, \ref{fdSactw}). The result is the simple expression:
\be
	e^{i\tau M_{n0}}|\,\vec p\,\rangle =e^{-(i\mu-{n-1\over 2})\tau}|\,e^\tau \vec p\,\rangle \,.
\ee
Finally we obtain the following expansion of the field operator: 
\be
	\Phi(\tau,\vec x)=\int d^{n-1}\vec p\, a^\dagger(\vec p)\,e^{i\vec p\cdot\vec x}e^{-(i\mu+{n-1\over 2})\tau}
	\langle\, e^{-\tau}\vec p\,|\Psi_0\rangle +{\rm h.c.}
\ee
The expressions (\ref{ffnn}) and (\ref{pp}) for the components of $|\Psi_0\rangle $ and $|\,\vec p\,\rangle $ allow
to calculate the amplitude $\langle\, \vec p\,|\Psi_0\rangle $ as
 \ba
	\langle\,\vec p\,\vert\Psi_0\rangle&=&
	\frac1{\sqrt2}\,\Gamma\left(i\mu-\frac{n-3}2\right)\,|\,\vec p\,|^{-i\mu}
	\nonumber\\
	&&\times\left[A\,\left\{\cos\left(\frac{n-3}2\pi\right)J_{-i\mu}
	(|\,\vec p\,|)+\sin\left(\frac{n-3}2\pi\right)Y_{-i\mu}(|\,\vec p\,|)\right\}
	+B\,J_{i\mu}(|\,\vec p\,|)\right]
	\nonumber\\
	&=&\frac1{\sqrt2}\,\Gamma\left(i\mu-\frac{n-3}2\right)\,|\,\vec p\,|^{-i\mu}\,
	\sinh\left(\pi\mu+i\pi\frac{n-1}2\right)\,
	\left[\,C\,H_{i\mu}^{(1)}(|\,\vec p\,|)-D\,H_{i\mu}^{(2)}(|\,\vec p\,|)\,\right].
	\nonumber\\
\ea
Notice that the $\tau\rightarrow-\infty$ limit is the same as the $|\,\vec p\,|\rightarrow
\infty$ limit. The BD vacuum which is characterized by positive conformal frequencies in the high momentum limit 
here coincides with the {\it in} vacuum with positive conformal frequencies in the $\tau\rightarrow-\infty$ limit.
The {\it out} vacuum is, as expected, the same one we got using the global sections.
The various vacua can be straightforwardly obtained using the asymptotic behavior of the Bessel
 functions \cite{bateman}:
\ba
	&&J_{i\mu}(z)\underset{z\to0}{\approx}
	\frac1{\Gamma(i\mu+1)}\left(\frac z2\right)^{i\mu}\,,
	\nonumber\\
	&&Y_{i\mu}(z)\underset{z\to0}{\approx}
	-iH_{i\mu}^{(1)}(z)\underset{z\to0}{\approx}
	iH_{i\mu}^{(2)}(z)\underset{z\to0}{\approx}
	-\frac{\Gamma(i\mu)}{\pi}\left(\frac z2\right)^{-i\mu}\,,
	\nonumber\\
	&&H_{i\mu}^{(1)}(z)\underset{z\to\infty}{\approx}
	\sqrt{\frac2{\pi z}}e^{\frac{\pi\mu}2}e^{i\left(z-\frac\pi4\right)}\,,
	\quad
	H_{i\mu}^{(2)}(z)\underset{z\to\infty}{\approx}
	\sqrt{\frac2{\pi z}}e^{-\frac{\pi\mu}2}e^{-i\left(z-\frac\pi4\right)}\,.
\ea

\end{document}